# Chatbots im Schulunterricht: Wir testen das Fobizz-Tool zur automatischen Bewertung von Hausaufgaben

Rainer Mühlhoff[1] ; Marte Henningsen[2]
Kontakt: rainer.muehlhoff .a.t. uni-osnabrueck.de

### Abstract (Deutsch)

Die vorliegende Studie untersucht das KI-gestützte Korrekturtool „KI-Korrekturhilfe“ des Unternehmens Fobizz, das Lehrkräften Unterstützung bei der Bewertung und Rückmeldung von Schülerarbeiten bieten soll. Im gesellschaftlichen Kontext eines überlasteten Bildungssystems und wachsender Erwartungen an den Einsatz von künstlicher Intelligenz zur Lösung dieser Probleme analysiert die Untersuchung die funktionale Eignung des Tools anhand von zwei Testreihen. Dabei zeigen die Ergebnisse erhebliche Defizite: Die numerischen Bewertungen und qualitativen Rückmeldungen des Tools hängen häufig vom Zufall ab und verbessern sich nicht durch die Einarbeitung der Verbesserungsvorschläge des KI-Tools. Eine Bestbewertung ist nur mit Texten erreichbar, die von ChatGPT geschrieben sind. Falschbehauptungen und Nonsense-Abgaben werden häufig nicht erkannt, und die Umsetzung einiger Bewertungskriterien ist unzuverlässig und intransparent. Da diese Mängel aus den fundamentalen Einschränkungen großer Sprachmodelle (LLMs) resultieren, sind grundlegende Verbesserungen dieses oder ähnlicher Tools nicht unmittelbar zu erwarten. Die Studie kritisiert den allgemeinen Trend, KI als schnelle Lösung für systemische Probleme im Bildungswesen einzusetzen. Sie kommt zu dem Schluss, dass die Vermarktung des Tools durch Fobizz als objektive und zeitsparende Lösung irreführend und unverantwortlich ist und mahnt zu systematischer Evaluation und fachdidaktischer Prüfung des Einsatzes von KI-Tools im Schulkontext.

### Abstract (English)

This study examines the AI-powered grading tool “AI Grading Assistant” by the German company Fobizz, designed to support teachers in evaluating and providing feedback on student assignments. Against the societal backdrop of an overburdened education system and rising expectations for artificial intelligence as a solution to these challenges, the investigation evaluates the tool's functional suitability through two test series. The results reveal significant shortcomings: The tool's numerical grades and qualitative feedback are often random and do not improve even when its suggestions are incorporated. The highest ratings are achievable only with texts generated by ChatGPT. False claims and nonsensical submissions frequently go undetected, while the implementation of some grading criteria is unreliable and opaque. Since these deficiencies stem from the inherent limitations of large language models (LLMs), fundamental improvements to this or similar tools are not immediately foreseeable. The study critiques the broader trend of adopting AI as a quick fix for systemic problems in education, concluding that Fobizz’s marketing of the tool as an objective and time-saving solution is misleading and irresponsible. Finally, the study calls for systematic evaluation and subject-specific pedagogical scrutiny of the use of AI tools in educational contexts.

---

[1] Prof. Dr. Rainer Mühlhoff, AG Ethik und kritische Theorien der KI, Institut für Kognitionswissenschaft, Universität Osnabrück, Wachsbleiche 27, 49090 Osnabrück, Deutschland.
[2] Marte Henningsen, Department of Philosophy, Faculty of Arts and Social Sciences, Maastricht University, Grote Gracht 90-92, 6211SZ Maastricht, The Netherlands.

# Executive Summary

Die vorliegende Studie untersucht das KI-gestützte Korrekturtool **„KI-Korrekturhilfe“** des Unternehmens **Fobizz**, das automatisiert Bewertungen und Rückmeldungen für Schülerarbeiten (Aufsätze, Hausaufgaben, Klassenarbeiten) generieren soll. Vor dem Hintergrund eines überlasteten Bildungssystems und wachsender Erwartungen an den Einsatz von KI als Problemlöser analysiert die Untersuchung die funktionale Eignung des Tools.

### Fokus

Die Studie konzentriert sich darauf, offensichtliche Mängel der Gebrauchstauglichkeit des Tools zu identifizieren, die allgemein als Schwächen angesehen werden, unabhängig von unterschiedlichen Meinungen über didaktische Konzepte und pädagogische Methoden.

### Methodik

Das Tool wurde anhand von zwei Testreihen geprüft. **Testreihe A** simulierte zehn Schülerarbeiten, die jeweils fünfmal unabhängig durch das Tool bewertet wurden. Dieses Design zielte darauf ab, die Konsistenz der numerischen Bewertung und des qualitativen Feedbacks zu untersuchen. Denn da das Tool im Hintergrund auf dem großen Sprachmodell (LLM) GPT-4 basiert, enthalten seine Ausgaben technisch bedingt Zufallselemente, was bedeutet, dass das Wiederholen desselben Bewertungsvorgangs zu unterschiedlichen Ergebnissen führen könnte. **Testreihe B** hat iterative Verbesserungen von Schülerarbeiten auf Basis des automatischen Feedbacks des Korrekturtools getestet. Wir wollten beobachten, ob sich die Gesamtbewertung durch Verbesserung der ausgewiesenen Fehler steigert.

### Hauptergebnisse

Die Studie zeigt mehrere erhebliche Mängel und Einschränkungen der Gebrauchstauglichkeit des Fobizz-Korrekturtools auf:

1. **Zufallsbedingte Volatilität von Bewertungen und Feedback**: Sowohl die vorgeschlagene Gesamtnote als auch das qualitative Feedback variierten erheblich zwischen verschiedenen Bewertungsdurchläufen derselben Abgabe. Diese Volatilität stellt ein ernstes Problem dar, da Lehrkräfte, die sich auf das Tool verlassen, unbemerkt quasi “ausgewürfelte” und potenziell ungerechte Noten und Rückmeldungen vergeben könnten.

2. **Unzuverlässige Erkennung von sachlichen Fehlern und Nonsense-Abgaben:** Das Korrekturtool konnte sachliche Falschbehauptungen sowie unsinnige oder arbeitsverweigernde Abgeben nicht zuverlässig erkennen.

3. **Inkonsistente Umsetzung von Bewertungskriterien**: Das Korrekturtool erlaubt es Lehrkräften, benutzerdefinierte Bewertungskriterien einzugeben. Die Studie fand jedoch heraus, dass die Qualität der Umsetzung dieser Kriterien stark variiert, was jedoch nicht transparent wird. Beispielsweise hatte das Tool Schwierigkeiten, Einreichungen konsistent auf Basis der Wortanzahl zu bewerten, und konnte KI-generierte Texte nicht erkennen – suggeriert in beiden Fällen aber, diese Kriterien zu beherrschen.

4. **Inkonsistentes Feedback**: Die Forschenden stellten mehrere Inkonsistenzen im generierten Feedback fest, darunter uneinheitliche Terminologie für Fehlerkategorien, das Auflisten von Fehlern, die nicht existierten, sowie widersprüchliche Angaben der Gesamtnote innerhalb desselben Feedbackdokuments.

5. **Umsetzung des Feedbacks führt nicht zu Verbesserungen:** Die vorgeschlagene Gesamtnote steigt bei schrittweiser Umsetzung der Verbesserungsvorschläge des Korrekturtools nicht. Die Note schwankt bei sukzessiven Verbesserungsläufen um die ursprüngliche Bewertung, verschlechtert sich in den ersten Iterationen sogar und bleibt im Durchschnitt von 7–12 Iterationen leicht unter der ursprünglichen Bewertung. Auch das qualitative Feedback folgt diesem Muster.

6. **Bestnote nur durch ChatGPT möglich (Täuschung):** Selbst mit vollständiger Umsetzung der Verbesserungsvorschläge war es nicht möglich, eine “perfekte” – d.h. nicht mehr beanstandete – Einreichung vorzulegen. Eine nahezu perfekte Bewertung gelang nur durch Überarbeitung der Lösung mit ChatGPT, was Schüler:innen signalisiert, dass sie für eine Bestnote auf KI-Unterstützung zurückgreifen müssen.

### Schlussfolgerungen

Das Tool weist grundlegende Defizite auf, von denen die Studie mehrere als “fatale Gebrauchshindernisse” klassifiziert (siehe Tabelle #tab:D:gravität). Es wird darauf hingewiesen, dass die meisten der beobachteten Mängel auf die inhärenten technischen Eigenschaften und Limitationen großer Sprachmodelle (LLMs) zurückzuführen sind. Aus diesen Gründen ist eine schnelle technische Lösung der Mängel nicht zu erwarten.

Die Ergebnisse legen nahe, dass das Tool für den Schulalltag ungeeignet ist. Die beobachteten Mängel werfen erhebliche Fragen hinsichtlich der Zuverlässigkeit und Fairness des Tools auf. Zudem kritisiert die Studie die unverantwortliche und irreführende Bewerbung und Vermarktung des Tools durch Fobizz.

### Empfehlungen

Die Studie fordert:

- Kein Erwerb von Flächenlizenzen für ungetestete KI-Tools im Schulkontext. Etablierung systematischer Evaluations- und Akkreditierungsprozesse für KI in der Bildung.
- Kritische Prüfung der didaktischen Eignung solcher Technologien.
- Lehrkräfte mit umfassendem technischem Hintergrundwissen über LLMs ausstatten.
- Politische Maßnahmen zur Verbesserung des Bildungssystems anstelle einer Fokussierung auf Automatisierung und KI zur Lösung struktureller Defizite.

Die “KI-Korrekturhilfe” von Fobizz erfüllt weder die technischen noch die didaktischen Mindestanforderungen für den praktischen Einsatz im Schulwesen und sollte nicht im Schulalltag verwendet werden.

# Inhalt

# I. Einleitung

## 1. Der Ruf nach mehr KI im Schulunterricht

Empirische Erhebungen und öffentliche Berichterstattung beschreiben seit Jahren einen besorgniserregenden Zustand des deutschen Bildungssystems. In einer Statista-Studie von 2024 geben nur 30% der befragten Bürger:innen an, mit dem Schulsystem "zufrieden" oder "sehr zufrieden" zu sein.[3] Bei einer Befragung der Robert Bosch Stiftung gaben knapp 70% der Lehrkräfte an, dass sie sich mindestens einmal die Woche durch ihre Arbeit erschöpft fühlten, und 43% gaben an, sie seien mindestens einmal die Woche bereits vor dem Beginn des Schultags wieder müde.[4] Die Autor:innen des Bildungsberichts 2024 sprechen davon, dass das System "am Anschlag arbeitet".[5] In wirtschaftlicher Hinsicht moniert das ifo-Zentrum für Bildungsökonomik, dass das Schulsystem unterfinanziert sei und an einem Investitionsrückstand leide.[6] In der Politik der aktuellen Legislatur wurden diese Befunde in den manischen Ruf nach "mehr Tempo" bei der Digitalisierung übersetzt – ein Argument, das nach der Corona-Pandemie häufig vorgebracht wird.[7] Dabei verweisen die meisten Studien als Ursache der Krise des Bildungssystems auf Lehrkräftemangel, Überarbeitung der vorhandenen Lehrkräfte, die insgesamt schlechten Arbeitsbedingungen an Schulen und mangelnde Wertschätzung des Berufs.

Spätestens seit der Markteinführung von ChatGPT, dem beliebten Chatbot-System des US-amerikanischen Unternehmens OpenAI, das seit Herbst 2022 verfügbar ist, wird "künstliche Intelligenz" (KI) auf der Basis großer Sprachmodelle (LLMs, large language models) als ein wesentlicher Baustein digitaler Lösungen für das Bildungsproblem gehandelt. Basierend auf den spezifischen (neuen) Möglichkeiten, die generative LLM-Systeme bieten, wird von zahlreichen kleineren Startups und großen Unternehmen (z.B. Microsoft) eine neue Generation von KI-Tools für den Bildungsbereich angeboten. Bei diesen Tools steht die text- und sprachbasierte Interaktion im Vordergrund: sei es zwischen Schüler:in und Computersystem, Lehrkraft und Computersystem, oder Schüler:in und Lehrkraft vermittelt durch KI-basierte Computersysteme. Während zum Beispiel in London derzeit eine ganze Klasse einer Privatschule "lehrerlos" in das neue Schuljahr startet und ausschließlich "von einer Kombination aus KI-Plattformen und Virtual Reality Brillen unterrichtet wird",[8] nehmen wir auch im Deutschen Kontext einen zwar weniger auf Vollautomatisierung abstellenden, dennoch aber immer lauter werdenden Ruf nach mehr KI-Technologie im Schulkontext wahr. So beobachten wir als wichtigste Anwendungen dieser Technologie im deutschsprachigen Diskurs vor allem die folgenden Trends:

---

[3] https://de.statista.com/statistik/daten/studie/159863/umfrage/zufriedenheit-mit-dem-schulsystem/

[4] https://www.bosch-stiftung.de/sites/default/files/documents/2022-06/41546_f22.0127_text_Schulbarometer_Gesundheit.pdf ; https://deutsches-schulportal.de/bildungswesen/deutsches-schulbarometer/ ; https://www.bosch-stiftung.de/sites/default/files/documents/2024-04/Schulbarometer_Lehrkraefte_2024_FORSCHUNGSBERICHT.pdf

[5] https://www.zdf.de/nachrichten/politik/deutschland/nationaler-bildungsbericht-probleme-schule-deutschland-100.html ; https://www.kmk.org/de/themen/bildungsberichterstattung/bildungsbericht-2024.html

[6] https://www.rnd.de/wirtschaft/ifo-umfrage-zum-schulsystem-in-deutschland-zufriedenheit-auf-tiefstand-EIF5A5WA7FKKJNHFFJ7CUUJS5Q.html

[7] https://www.tagesschau.de/inland/innenpolitik/bildung-stark-watzinger-101.html

[8] https://www.avinteractive.com/news/virtual-augmented-mixed/students-in-uks-first-teacherless-classroom-taught-by-ai-02-09-2024/ ; https://www.businessinsider.com/chatgpt-ai-tools-replace-teachers-high-school-students-learning-education-2024-8?op=1

1. **Automatisierung von Korrektur, Rückmeldungen und Bewertung von Lernleistungen (Aufsätze, Klassenarbeiten, …).** Sowohl in der Variante denkbar, dass Leher:innen ein solches Tool unterstützend nutzen, als auch dass Schüler:innen direkt mit einem KI-System korrespondieren, um Feedback für Lernleistungen zu erhalten.

2. **Chatbots als "Lerntutoren":** In schriftlicher Interaktion zwischen Schüler:in und Computersystem werden hier z.B. Lerninhalte schrittweise erklärt, die Bearbeitung von (Haus-)Aufgaben betreut oder allgemeine Mentoringfragen diskutiert.

3. **Chatbots als "Sparring Partner" für Lehrkräfte:** Lehrer:innen benutzen Chatbots, um Ideen und Anregungen für Unterrichtsgestaltung, Planung, und sonstige Aufgaben des Lehrerberufs zu erhalten.

4. **Simulierte Konversationen mit Avataren:** Chatbots simulieren (vorgeblich) das Gesprächsverhalten, die inhaltlichen Ansichten oder gar den "Charakter"[9] einer bekannten Persönlichkeit wie Sokrates oder Sophie Scholl; Interaktionen mit einem solchen Bot werden als Lernmaterialien verwendet.

5. **Automatisierte Herstellung von Lernmaterialien und Unterrichtskonzepten.** Hier wird oft die Dimension der Individualisierung betont: LLMs werden verwendet, um dem Lernstand einzelner Schüler:innen oder ganzer Schulklassen angepasste Lernmaterialien und -konzepte zu erstellen.

Ein häufig vorgebrachtes Argument für die Anschaffung von LLM-basierten KI-Tools im Bildungskontext lautet, dass in einer Situation des Lehrkräftemangels KI-Systeme einen Teil der Arbeitslast übernehmen könnten, um den Lehrkräften mehr Zeit für das Wesentliche ihrer Arbeit – den Kontakt mit den Schüler:innen – zu ermöglichen.[10] Eine Variante dieses Arguments betont, dass die Automatisierung zusätzlich eine Individualisierung der pädagogischen Betreuung einzelner Schüler:innen ermögliche, die den Lehrkräften aktuell angesichts großer Klassenstärken und steigender "Heterogenität der Schüler:innen" kaum zuzumuten sei.[11] Gerade bei automatisierten Korrekturen und Mentoring-Diensten wird des Weiteren argumentiert, ein Vorteil der Automatisierung bestehe in der Vermeidung eventueller Biases und Vorurteilen der Lehrkräfte, da das Computersystem datenbasiert und somit objektiv arbeite.[12] In dieser Argumentationsweise schlägt sich das seit den 1990er Jahren wissenschaftlich dokumentierte psychologische Phänomen des "automation bias" (dt. etwa "Automatisierungsvoreingenommenheit") nieder, nach dem Menschen die Tendenz haben, automatisierten Entscheidungen oder Einschätzungen mehr zu vertrauen, als menschlichen Entscheidungen – selbst angesichts widersprechender Faktenlage.[13]

Neben diesen im öffentlichen Diskurs vorgebrachten Argumenten für die Einführung LLM-basierter KI-Tools im Bildungskontext beobachten wir in unserer Arbeit an Schulen und mit Lehrer:innen, dass an vielen Stellen im Schulsystem in den letzten Jahren eine Dynamik der Eigeninitiative zu verzeichnen war: Das Ausprobieren oder sogar Anschaffen solcher Tools auf individueller oder Schulebene scheint nicht selten dem Engagement und der Vorreiterfunktion einiger, tendenziell technikbegeisterter (dazu häufig jüngerer und männlicher) Lehrkräfte zu entspringen, die an ihren Schulen und unter ihren Kolleg:innen entsprechende Aufmerksamkeit auf das Thema KI als “Gamechanger” im Bildungskontext richten und Trends in Gang setzen. Dieses von unten kommende Engagement wird seit kürzerer Zeit komplementiert durch das Interesse einiger Bundesländer und Schulträger, entsprechende Softwaresysteme von zentraler Stelle anzuschaffen oder in Pilotprojekten zu testen, teilweise auch Kompetenzen in diesen Bereichen auf landeseigenen “Schulplattformen” und “Bildungsservern” zu bündeln und zur Verfügung zu stellen.[14]

[9] Mit dieser vollmundigen Ankündigung bewirbt Fobizz seinen "AI Character Chat", https://tools.fobizz.com/.

[10] https://deutsches-schulportal.de/bildungswesen/weniger-teilzeit-groessere-klassen-umgang-mit-lehrermangel-empfehlungen-staendige-wissenschaftliche-kommission/ ; https://www.kmk.org/fileadmin/veroeffentlichungen_beschluesse/2024/2024_10_10-Handlungsempfehlung-KI.pdf ; https://www.bpb.de/shop/zeitschriften/apuz/kuenstliche-intelligenz-2023/541500/ki-in-der-schule/ ; https://www.news4teachers.de/2023/06/mckinsey-studie-kuenstliche-intelligenz-wird-lehrberufe-am-staerksten-veraendern/

[11] https://www.bpb.de/shop/zeitschriften/apuz/kuenstliche-intelligenz-2023/541500/ki-in-der-schule/ ; https://www.kmk.org/fileadmin/veroeffentlichungen_beschluesse/2024/2024_10_10-Handlungsempfehlung-KI.pdf ; https://www.cornelsen.de/magazin/beitraege/die-potenziale-von-ki-in-der-bildung

[12] https://deutsches-schulportal.de/bildungswesen/weniger-teilzeit-groessere-klassen-umgang-mit-lehrermangel-empfehlungen-staendige-wissenschaftliche-kommission/

[13] Skitka, L.J., Mosier, K.L. and Burdick, M. (1999) 'Does automation bias decision-making?' *International Journal of Human-Computer Studies*, 51(5): 991–1006, doi: 10.1006/ijhc.1999.0252.

So zeigt sich aktuell im deutschlandweiten Vergleich eine heterogene Gesamtsituation, in der seit dem Jahr 2023 einige Bundesländer Lizenzverträge mit Unternehmen geschlossen haben, die LLM-basierte KI-Tools für alle Lehrkräfte oder Schüler:innen anbieten, während sich andere Bundesländer in Testphasen oder Pilotprojekten befinden und wiederum andere keine erkennbaren Aktivitäten zeigen. Unsere Rechercheergebnisse und Kenntnisse hierzu sind in Tabelle #tab:E:bundesländer zusammengefasst; wir haben außerdem fast alle Bundesländer zum Stand der Dinge um Stellungnahmen gebeten.

| | |
|---|---|
| Baden-Württemberg | Seit Februar 2024 im Pilotversuch mit landeseigenem Tool ‘fAIrChat’, welches auch eine Schnittstelle zu ChatGPT bietet. “Ziel ist es, zum Herbst 2024 eine gute Entscheidungsgrundlage zu haben, inwieweit fAIrChat in der Fläche zur Verfügung gestellt werden soll.” |
| Bayern | Seit 2022 Modellversuch “ki@school” mit 19 Schulen. Schulen des Landes können Anträge auf “Medien- und KI-Budget” stellen und davon auf Schulebene Lizenzen und Software anschaffen.[15] |
| Berlin | Verkündet am 22.10.2024 über den Instagram Account “senbildjugfam” (Berliner Bildungsministerium): “Als eines der ersten Bundesländer bietet Berlin seinen Lehrkräften jetzt ein datenschutzkonformes KI-Tool an. Mit **Copilot,** dem KI-Assistenten von Microsoft wird alles digitaler.”[16] |
| Brandenburg | Seit Sommer 2024 landeseigene Plattform “jwd” (jetzt wird’s digital) mit einem Informationsangebot, aber ohne KI-Tools oder -Lizenzen. |
| Bremen | Unklare Informationslage; im Februar 2023 scheint Bremen eine Pilotphase mit **Fobizz** gestartet zu haben.[17] |
| Hamburg | Nichts bekannt. |
| Hessen | Pilotprojekt “KI4S’cool” mit 25 Schulen zum individualisierten Lernen mit KI in der gymnasialen Oberstufe. Keine Lizenz-Verträge für KI-Tools bekannt.[18] |
| Mecklenburg-Vorpommern | Landeslizenz mit **Fobizz** (läuft vorerst bis Ende 2024). Seit April 2020 Landeslizenz für die Fobizz-Fortbildungen, seit September 2023 auch für |

[14] https://deutsches-schulportal.de/unterricht/fobizz-schulki-und-co-welche-ki-tools-koennen-schulen-nutzen/
[15] https://www.bildunginbayern.de/berufliche-bildung/kischool/ ; https://www.km.bayern.de/gestalten/foerderprogramme/medien-und-ki-budget
[16] https://www.tagesspiegel.de/berlin/mit-copilot-von-microsoft-berliner-lehrer-bereiten-unterricht-jetzt-mit-kunstlicher-intelligenz-vor-12552278.html
[17] https://www.vidis.schule/pilotphase-bremen-fobizz/
[18] https://kultus.hessen.de/presse/mit-kuenstlicher-intelligenz-unterricht-individueller-zielgerichteter-und-effektiver-gestalten

| | |
|---|---|
| | die KI-Tools.[19] |
| Niedersachsen | Nichts bekannt. |
| Nordrhein-Westfalen | Pilotversuch “KIMADU” (KI im Mathematik- und Deutschunterricht) im Jahr 2025; seit Feb. 2023: Handlungsleitfaden “Umgang mit textgenerierenden KI-Systemen”.[20] |
| Rheinland-Pfalz | **Fobizz**-Landeslizenz, Laufzeit Dezember 2023 bis Juli 2025, Kosten: 1.77 Mio € netto.[21] |
| Saarland | Unklare Informationslage. Im Jahr 2023 scheint das Saarland eine Pilotphase mit **Fobizz** gestartet zu haben. Außerdem scheint das Saarland Prüfungen mittels KI einzuführen.[22] |
| Sachsen | Landeslizenz von **Fobizz** für die Lehrerfortbildung; außerdem landeseigenes KI Tool “KAI” mit Schnittstelle zu ChatGPT für alle Lehrkräften in Sachsen.[23] |
| Sachsen-Anhalt | Landeseigenes Tool “emuKI GPT-4”, welches auf ChatGPT beruht, landesweit kostenlos nutzbar aber verpflichtende Onlinefortbildung vorher.[24] |
| Schleswig-Holstein | Pilotprojekt mit 66 Schulen, Handreichung “KI@Schule”.[25] |
| Thüringen | Nichts bekannt. |

**Tab. #tab:E:bundesländer:** Unser Recherche- und Kenntnisstand zur Anschaffung von LLM-basierten KI-System für Lehrkräfte und/oder Schüler:innen in den Bundesländern. rot = es gibt zentrale Lizenzverträge oder landeseigene Software-Tools; gelb = Testphasen, Pilotprojekte, Sondierungen. blau = uns sind keine Informationen bekannt.

**Haben Sie Hinweise und Informationen zu Ihrem Bundesland?** Unterstützen Sie unsere Forschung, kontaktieren Sie uns gerne unter **KIundSchule@ethikderki.de**!

Insgesamt befindet sich die Bildungslandschaft in Deutschland in Bezug auf die Anschaffung von LLM-basierten KI-Tools zur Zeit in einer fortgeschrittenen Experimentierphase, in der zunehmend durch öffentliche Gelder (potenziell experimentelle und noch schlecht untersuchte) Softwaretools angeschafft und im Schulalltag eingesetzt oder getestet werden. Wir bezeichnen diese Phase als “experimentell”, weil sich bisher weder ein Kanon besonders geeigneter Softwaretools herauskristallisiert hat, noch didaktischer Konsens über den tatsächlichen Nutzen dieser Tools besteht, noch die Qualität der Tools im Bildungskontext systematisch evaluiert und wissenschaftlich beschrieben worden ist. Schließlich beobachten wir, dass in dieser Gemengelage die optimistischen Stimmen in Bezug auf den Einsatz von KI-Technologie den Ton angeben, während eine verantwortungsbewusste didaktische Auseinandersetzung damit vergleichsweise weniger weit fortgeschritten ist und kritische Stimmen weniger gehört werden.

[19] https://fobizz.com/lehrerfortbildung-mecklenburg-vorpommern/ ; https://www.bildung-mv.de/aktuell/2023/fobizz-fuer-lehrkraefte-in-mv-verlaengert/

[20] https://www.schulministerium.nrw/pilotprojekt-kimadu ; https://www.lernen-digital.nrw/ki-arbeitshilfen ; https://www.schulministerium.nrw/system/files/media/document/file/handlungsleitfaden_ki_msb_nrw_230223.pdf

[21] Persönliche Korrespondenz mit dem Ministerium für Bildung, Rheinland-Pfalz.

[22] https://www.vidis.schule/saarland-abschluss-pilotphase/ ; https://www.sr.de/sr/home/nachrichten/panorama/neue_regeln_zur_leistungsbewertung_an_saar-schulen_100.html

[23] https://www.unterstuetzung-sachsen.de/angebot/detail.php?aid=922 ; https://www.bildung.sachsen.de/blog/index.php/2024/10/23/kuenstliche-intelligenz-in-der-schule/

[24] https://www.bildung-lsa.de/digitale_dienste/emuki_gpt_4.html ; https://mb.sachsen-anhalt.de/details/lehrkraefte-erhalten-kostenlosen-zugang-zu-innovativen-ki-werkzeugen

[25] https://www.schleswig-holstein.de/DE/landesregierung/themen/bildung-hochschulen/digitale-schule/Lernen/ki_schule/KI_Schule

## 2. Der deutsche Marktführer “Fobizz”

Die vorliegende Studie leistet einen pointierten Beitrag zu dieser Auseinandersetzung, indem sie ein konkretes KI-Tool für Lehrkräfte des Hamburger Unternehmens Fobizz unter die Lupe nimmt und hinsichtlich seiner Qualität und Gebrauchstauglichkeit qualitativ einordnet. Fobizz (https://fobizz.com/) ist ein kleines Unternehmen, das laut Eigendarstellung ca. 40 Mitarbeitenden beschäftigt und sich selbst als “ein Startup aus dem EdTech Bereich” bezeichnet.[26] Gegründet wurde es 2018 zunächst als digitale Plattform für Lehrkräfte-Fortbildungen;[27] die Sparte “digitale Tools und KI für Lehrkräfte und Schulen”, die für unsere Untersuchung im Zentrum steht, kam im Laufe der Jahre hinzu, vermutlich nach der Markteinführung von ChatGPT im Herbst 2022, da viele der von Fobizz angebotenen KI-Tools im Hintergrund auf ChatGPT und anderen neueren generative KI Systemen großer internationaler Unternehmen beruhen.[28]

Im Kontext unserer Studie fällt die Wahl deshalb auf Fobizz, weil das Unternehmen in den letzten Jahren unter den deutschen Firmen für KI-Tools im Schulbereich einer der Marktführer geworden zu sein scheint.[29] Fobizz bietet eine differenzierte Palette von Nutzungslizenzen für sein Software-Bündel bestehend aus digitalen Fortbildungen, KI-Tools und Unterrichtsmaterialien an, unter anderem für einzelne Lehrkräfte, Schulen, Schulträger, Bundesländer, Medienzentren sowie Studierende und Referendar:innen (https://fobizz.com/preise/). Mit Stand November 2024 ist uns bekannt, dass mindestens drei Bundesländer – Rheinland-Pfalz, Mecklenburg-Vorpommern und Sachsen – mit Fobizz Landeslizenzen für alle Lehrkräfte des Landes abgeschlossen haben,[30] weitere zwei Bundesländer (Bremen und das Saarland) kooperieren oder kooperierten auf andere Weise mit Fobizz (u.a. im Rahmen von Modellversuchen im Jahr 2023)[31]; weitere Länder oder Schulträger prüfen aktuell den Erwerb einer Fobizz-Lizenz für ihre Lehrkräfte. Darüber hinaus können einzelne Schulen oder Lehrkräfte direkt Lizenzen bei Fobizz erwerben. Insgesamt wirbt Fobizz damit, dass mehr als 7.500 Schulen und 500.000 Lehrkräften in Deutschland, Österreich und der Schweiz die Produkte des Unternehmens nutzen.

Neben der Präsenz in Beschaffungsdiskussionen zeichnet sich Fobizz durch eine engagierte Werbestrategie aus. “Lass unsere KI-Assistenz dir dabei helfen, deinen Unterricht zu gestalten und deine Arbeit zu organisieren.”[32] oder “Spare wertvolle Zeit beim Korrigieren durch KI-gestützte Fehlerkorrektur und Bewertungsvorschläge.”[33] – so lauten die Beschreibungen verschiedener Tools auf der Website. Auf der Videoplattform YouTube sowie auf der Firmeneigenen Website finden sich zahlreiche als Informationsmaterial oder Fortbildungen stilisierte Videos, die zugleich Werbematerial für die Hauseigenen Softwaretools zu sein scheinen. Im Oktober 2024 verkündet Fobizz eine Kooperation mit dem Format “ZDF goes Schule” Zweiten Deutschen Fernsehen.[34] An Lehrer:innen, die sich auf der Fobizz-Website registriert haben (z.B. zur Teilnahme an einer Online-Schulung oder zur Nutzung der Softwaretools), werden wöchentliche Werbemails verschickt mit Betreffzeilen wie: “Lass die KI für dich arbeiten” (17.07.2024) oder “So verändert KI deinen Unterricht” (20.10.2024). Mittels bunter Emojis und niedlicher KI-Avataren werden die neuesten Produkte sowie Rabatt- und

[26] https://fobizz.com/was-ist-fobizz/
[27] https://fobizz.com/fobizz_wird_vier_jahre_alt/
[28] http://fobizz.w4p.tech/ki-assistenz-fobizz-tools/#toggle-id-6
[29] “Die am häufigsten genutzte und in allen Bundesländern verfügbare KI-Plattform ist fobizz” – https://deutsches-schulportal.de/unterricht/fobizz-schulki-und-co-welche-ki-tools-koennen-schulen-nutzen/
[30] https://fobizz.com/angebote-fuer-bundeslaender/
[31] Bremen: https://www.vidis.schule/pilotphase-bremen-fobizz/ ; Saarland: https://www.vidis.schule/saarland-abschluss-pilotphase/
[32] https://tools.fobizz.com/ai/chats/info
[33] https://tools.fobizz.com/ai/feedbacks/info
[34] https://www.facebook.com/story.php?story_fbid=1342464943629711&id=100035986545012&_rdr ; https://www.instagram.com/fobizz_com/p/DBJfu6HswUv/

Werbeaktionen der Firma angepriesen. Pünktlich zum Start der Sommerferien bewerben die E-Mails einen Mindfulness KI-Assistenten für eine “starke mentale Gesundheit” (21.07.2024), während ab Ende Oktober die Anmeldung zum Fobizz-Adventskalender in sechs verschiedenen Werbemails empfohlen wird. Mitarbeiter:innen von Fobizz sind außerdem auf zahlreichen Fachtagungen und Fortbildungsveranstaltungen für Lehrkräfte an Schulen mit persönlichen Vorträgen präsent.[35]

## 3. Scope der Untersuchung: Die “KI-Korrekturhilfe” von Fobizz

Den Fokus der vorliegenden Studie bildet ein exemplarisches KI-Tool der Firma Fobizz – der “AI Grading Assistant”, deutsch: “KI-Korrekturhilfe”, fortan als “Korrekturtool” bezeichnet. Es handelt sich dabei um einen ChatGPT-basierten, vorkonfigurierten Bot, der im Portal der Firma zur Verwendung durch Lehrkräfte angeboten und mit den Worten “[e]rhalte Bewertungsvorschläge und Hilfe bei Fehlerkorrekturen” beworben wird. Der Zweck des Bots ist es, textbasierte Arbeiten der Schüler:innen, z.B. Aufsätze oder Klassenarbeiten, mit strukturiertem Feedback und einem Bewertungsvorschlag zu versehen.

Unsere Wahl fiel aus mehreren Gründen auf genau dieses Angebot. Zum einen gehören Bewertung und Feedback seit jeher zu den Kernaufgaben des Lehrerberufs. Bewerten und insbesondere das Erstellen ausführlicher und angemessener Rückmeldungen kann extrem zeitaufwändig sein, während sich in dieser Kommunikation zugleich ein wichtiger Teil der pädagogischen und didaktischen Arbeit vollzieht. Für Schüler:innen können Bewertungen lebenswegentscheidend sein; die faire, transparente und vergleichbare Erstellung von Bewertungen, ihre angemessene Begründung und die Begleitung von konstruktivem, zur Verbesserung anleitenden Rückmeldungen sind entscheidend für den Lernerfolg. In dieser Situation trifft das an Lehrer:innen offensiv kommunizierte Wertversprechen, “Spare wertvolle Zeit beim Korrigieren durch KI-gestützte Fehlerkorrektur und Bewertungsvorschläge”[36], einen neuralgischen Punkt der systematischen Überlastung von Lehrkräften, die in dieser Überlastung mitunter Einbußen der pädagogischen Qualität ihrer Arbeit hinnehmen müssen und nun eine vermeintliche technische Lösung für dieses Problem angepriesen bekommen. Zugleich ist diese Beschreibung des Tools recht vollmundig, da sie suggeriert, eine Automatisierung der pädagogischen Arbeit des Beurteilens und Bewertens sei technisch möglich und moralisch vertretbar.

Dass diese Automatisierung technisch möglich ist, bezweifeln wir vor dem Hintergrund der Ergebnisse unserer Studie. Wir beschränken uns dabei auf eine rein funktionale Prüfung anhand zweier Testreihen, durch die wir eine Reihe gravierender Gebrauchshindernisse der Fobizz “KI-Korrekturhilfe” aufdecken (siehe Abschnitt II und folgende). Auf die moralische Dimension gehen wir in dieser Untersuchung nicht ein, noch weniger auf die rechtliche, zu der wir nur anmerken, dass es etwa in der Handreichung “KI@Schule” des Ministeriums für Allgemeine und Berufliche Bildung, Wissenschaft, Forschung und Kultur in Schleswig-Holstein (2023) heißt:

> **Verwendung von ChatGPT zur Bewertung von Schülerleistungen**
>
> Lehrkräfte haben die Leistungen der Schülerinnen und Schüler selbstständig aufgrund eigener Wahrnehmung zu bewerten. In § 16 Absatz 2 SchulG heißt es insoweit:
>
> *„Die beteiligten Lehrkräfte (…) bewerten die Leistungen der Schülerinnen und Schüler in pädagogischer Verantwortung.“ (vgl. z. B. auch § 13 Absatz 4 GemVO, §§ 19 Absatz 1, 29 Absatz 1 OAPVO).*
>
> Die jeweilige Leistungsbewertung ist für die Schülerin oder den Schüler insbesondere in Bezug auf die für die weitere Schullaufbahn und Abschlüsse maßgeblichen Zeugnisse von erheblicher Bedeutung. Die Bewertung von (Prüfungs-)Leistungen durch KI ist demnach unzulässig. Die Bewertung der Schülerleistung muss eine persönliche Eigenleistung der Lehrkraft sein.

[35] Siehe beispielsweise: https://appcamps.de/ki-summit-fuer-lehrkraefte/
[36] https://tools.fobizz.com/ai/feedbacks/info

## Aufbau und Verwendung des Korrekturtools

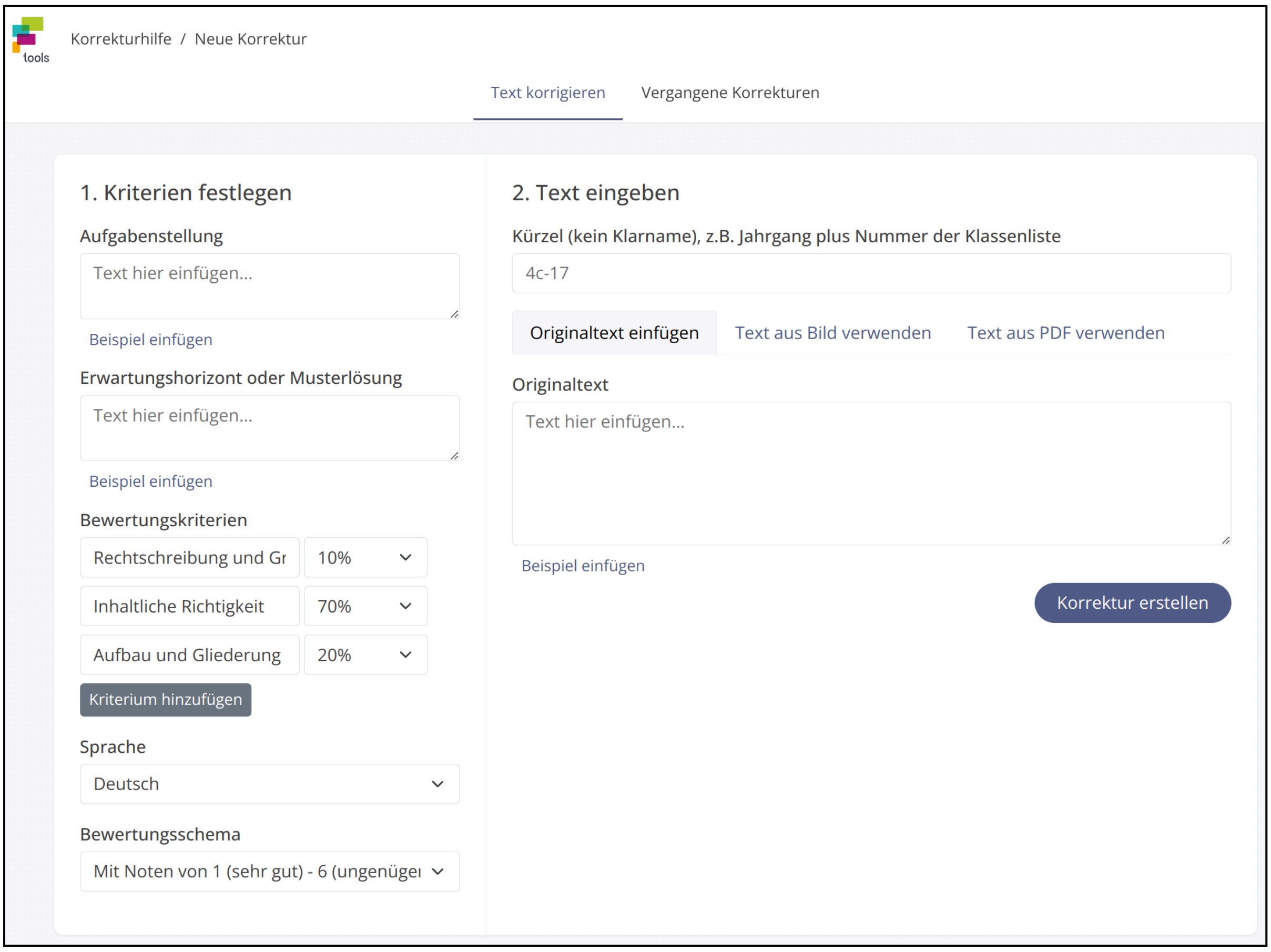

**Abb. #fig:E:korrekturtool-blank:** Nutzerinterface des Fobizz-Korrekturtools. Linke Spalte: Einstellung der Aufgabenstellung und Bewertungsmodalitäten. Rechte Spalte: Eingabe des zu korrigierenden Texts. Screenshot fobizz.com [28.11.2024].

Zur Verwendung muss der Fobizz “KI-Korrekturhilfe” muss der Bot anhand von 5 simplen Eingaben für die spezifische Aufgabenstellung eingerichtet werden (siehe linke Spalte in Abbildung #fig:E:korrekturtool-blank): (1) Ein Textfeld “Aufgabenstellung” dient der Eingabe der Aufgabenstellung, die auch die Schüler:innen erhalten haben; des Weiteren steht (2) ein Textfeld “Erwartungshorizont oder Musterlösung” zur Verfügung, (3) die Möglichkeit der freien Eingabe einer beliebig langen (oder kurzen) Liste von “Bewertungskriterien” samt relativer Gewichtung in Prozent, (4) ein Drop-Down Menü für die Auswahl der Sprache (Deutsch, Englisch, Französisch, Spanisch) und (5) ein Drop-Down Menü zur Auswahl des “Bewertungsschemas” (siehe Abbildung #fig:T:gradingscales).

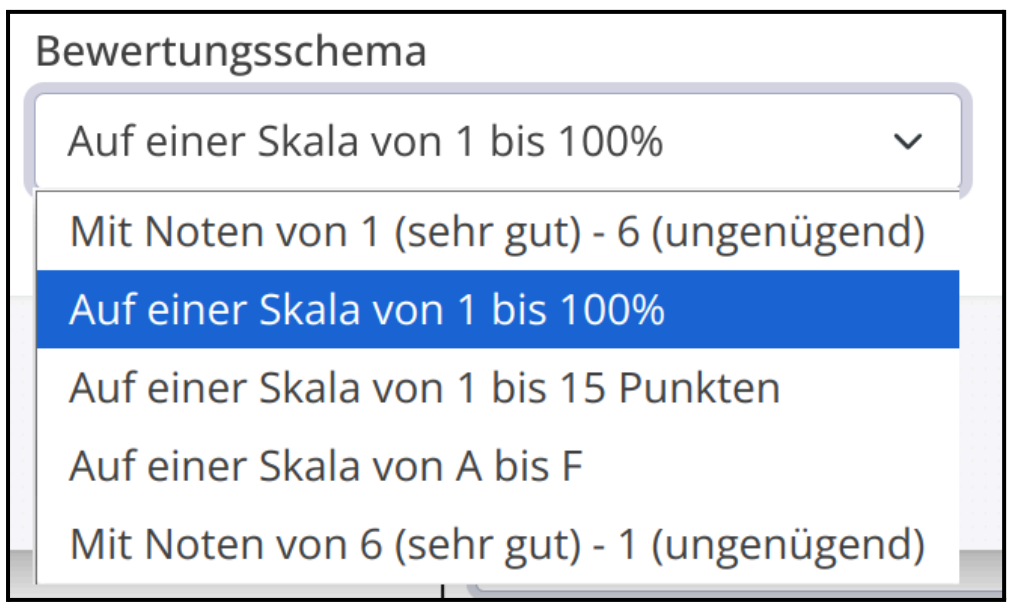

**Abb. #fig:T:gradingscales:** Dropdown-Menü der einstellbaren Bewertungsschemata. Screenshot fobizz.com [28.11.2024].

← **3 von 50** | Melissa 5.3 | 28.08.2024, 17:20 Uhr →

Originaltext

Sollte das Wahlalter auf 14 geändert werden?
Die Änderung des Wahlalters wird viel diskutiert. Im Folgenden möchte ich argumentieren, wieso ich eine Absenkung auf 14 für eine schlechte Idee halte.
Häufig ist die Forderung nach einer Änderung des Wahlalters verbunden mit einer Forderung nach mehr Mitbestimmung für junge Menschen. Doch zum einen gibt es viele andere Möglichkeiten der Mitbestimmung, z.B.

## Fehlerliste

| Fehler | Korrektur | Fehlerart |
|---|---|---|
| "wieso ich eine Absenkung auf 14 für eine schlechte Idee halte." | "warum ich eine Absenkung auf 14 Jahre für eine schlechte Idee halte." | Sprachstil |
| "z.B." | "zum Beispiel" | Abkürzung |
| "Das zeigt sich z.B. darin, dass sie häufiger extreme Parteien wählen." | "Das zeigt sich zum Beispiel darin, dass sie häufiger extreme Parteien wählen." | Abkürzung |
| "reif zu werden wenn man schließlich tatsächlich wählen darf." | "reif zu werden, wenn man schließlich tatsächlich wählen darf." | Kommafehler |

## Was du gut gemacht hast

- **Klare Argumentation:** Du hast sowohl ein Argument für als auch ein Argument gegen die Absenkung des Wahlalters auf 14 Jahre klar und verständlich dargestellt.
- **Logische Schlussfolgerung:** Deine Schlussfolgerung ist logisch und basiert auf den vorher genannten Argumenten.

## Was du verbessern kannst

- **Sprachstil und Formulierungen:** Verwende weniger umgangssprachliche Formulierungen und achte auf eine präzise Ausdrucksweise.
- **Vermeidung von Abkürzungen:** Vermeide Abkürzungen wie "z.B." und schreibe diese Begriffe aus, um die Lesbarkeit zu erhöhen.

## Detaillierte Bewertung

| Kategorie | Bewertung | Feedback |
|---|---|---|
| Rechtschreibung und Grammatik | 90% | Insgesamt sehr gut, nur wenige kleine Fehler wie fehlende Kommata und Abkürzungen. |
| Inhaltliche Richtigkeit | 95% | Die Argumente sind gut durchdacht und korrekt dargestellt. |
| Aufbau und Gliederung | 85% | Der Text ist gut strukturiert, könnte aber durch eine klarere Einleitung und ein stärkeres Fazit verbessert werden. |
| Nicht von KI geschrieben | 100% | Der Text wirkt authentisch und menschlich geschrieben. |
| Umfang 150 - 250 Wörter | 100% | Der Text liegt im geforderten Wortumfang. |

## Gesamtbewertung

Insgesamt hast du eine sehr gute Stellungnahme verfasst. Die Argumente sind klar und logisch, und der Text ist gut strukturiert. Achte in Zukunft auf eine präzisere Ausdrucksweise und vermeide Abkürzungen, um die Lesbarkeit weiter zu verbessern.

## Gesamtnote

93.25

Kopieren | PDF speichern | Bearbeiten

**Abb. #fig:E:bewertung:** Die Ausgabe ("Bewertung") des Korrekturtools für eine exemplarische Abgabe in der Ansicht im Webbrowser. Screenshot fobizz.com [28.11.2024].

Ist das Korrekturtool anhand dieser Eingaben für eine spezifische Aufgabenstellung konfiguriert worden, kann die Anwender:in die Arbeit einer Schüler:in in das Textfeld "Originaltext" kopieren (alternativ ein Bild oder eine PDF-Datei mit der Arbeit der Schüler:in hochladen) sowie ein Kürzel/Pseudonym für die jeweilige Schüler:in eingeben (das dient der Archivierung der Korrektur in der Liste vergangener Korrekturen), und schließlich auf den Button "Korrektur erstellen" drücken (siehe die rechte Spalte in Abbildung #fig:E:korrekturtool-blank).

Als Ergebnis erscheint dann eine textbasierte Ausgabe – fortan von uns als "Bewertung" bezeichnet –, die folgende Abschnitte enthält (siehe Abbildung #fig:E:bewertung): "Fehlerliste" (mit einer tabellarischen Aufstellung von "Fehler", "Korrektur"(-vorschlag) und "Fehlertyp"), "Was du gut gemacht hast" (Bulletpoint Liste), "Was du verbessern kannst" (Bulletpoint Liste) sowie "Bewertung" (erneut mit einer Tabelle, in der die einzelnen Bewertungskriterien angeführt, für jedes Kriterium eine Teilnote und ein "Feedback" angegeben werden). Schließlich folgt unter der Überschrift "Gesamtbewertung" ein textuelles Kurzfeedback und unter "Gesamtnote" eine numerische Gesamtbewertung, die dem in der Voreinstellung gewählten "Bewertungsschema" folgt. Diese Ausgabe kann man anhand von drei Buttons entweder in die Zwischenablage kopieren, als PDF Exportieren oder editieren.

# II. Untersuchungsziele und Methodik

Unser Ziel ist die qualitativen Untersuchung des Korrekturtools ("AI Grading Assistant") von Fobizz. Eine solche Untersuchung steht vor mindestens zwei besonderen Herausforderungen, einer prinzipiellen und einer technischen.

**Prinzipiell:** Die Produktion von Feedback und sogar Bewertungen auf Lernleistungen ist stets Gegenstand verfeinerter fachdidaktischer und pädagogischer Expertise, systematischer Ausbildung und Trainings, sowie der Kultivierung zwischenmenschlicher Sensibilitäten und Fähigkeiten. Die Beurteilung der Qualität von automatisch (oder auch nicht automatisch) produzierten Rückmeldungen und Bewertungen ist somit eine normativ hochgradig investierte und mitunter umstrittene Frage. Man wird sie nicht angehen können, ohne implizit oder explizit ein Verständnis davon zugrunde zu legen, was überhaupt "gutes" und "angemessenes" Feedback und eine "gute" und "angemessene" Bewertungspraxis ist. In diese Debatten möchten wir mit dieser Untersuchung jedoch nicht einsteigen – denn eine allgemeingültige und hinreichend leicht operationalisierbare Antwort hierauf zu finden erscheint uns vermessen und aussichtslos.

Wir beschränken uns deshalb darauf, nach offensichtlichen *Unzulänglichkeiten* des Korrekturtools zu suchen. Damit meinen wir Inkonsistenzen, die bei allen Differenzen um didaktische und pädagogische Methodologie mehr oder weniger *unumstritten als Mankos* aufgefasst werden. Wir untersuchen also *nicht*, ob das Korrekturtool *gutes* oder *angemessenes* Feedback produziert, sondern nur, ob es überhaupt die Mindestvoraussetzungen dafür erfüllt. Im Umkehrschluss heißt das nicht, dass wenn alle hier angeführten "Fehler" des Tools behoben wären, sein Einsatz automatisch empfehlenswert ist. Bei den im Folgenden angeführten Punkten handelt es sich nur um offensichtliche Missstände, durch die das Tool den notwendigen Mindeststandard für einen realen Einsatz verfehlt. Die Behebung dieser Missstände ist eine notwendige, aber keine hinreichende Voraussetzung, den Einsatz des Korrekturtools befürworten zu können.

**Technisch:** Das Korrekturtool von Fobizz basiert auf dem großen Sprachmodell GPT-4 von OpenAI, das auch in dem bekannten Chatbot "ChatGPT" zum Einsatz kommt.[37] Es liegt in den technischen Eigenschaften solcher Chatbotsysteme begründet, dass die "Antworten" auf "Prompts" randomisiert werden, also nicht-deterministisch durch den Prompt bestimmt sind. Das bedeutet: Mehrfache Wiederholung desselben Prompts führt zu mehr oder weniger stark unterschiedlichen Antworten, weil Zufallsfaktoren in die Verarbeitung einbezogen werden. Diese technische Eigenschaft, die von ChatGPT gut bekannt ist und leicht selbst reproduziert werden kann, führt zu dem methodologischen

[37] https://fobizz.com/die-fobizz-tools-fuer-schule-und-unterricht/ [aufgerufen 2024-11-29].

Problem, dass die Ergebnisse des Korrekturtools, also das Bewertungsdokument, das für eine konkrete Schülerabgabe produziert wird, nicht reproduzierbar ist – bei einem zweiten Bewertungsdurchlauf sieht das Dokument anders aus, inklusive inhaltlicher Abweichungen des Feedbacks und der Note. Wir dokumentieren deshalb die einzelnen Eingaben und Ausgaben unserer Untersuchung im Materialanhang. Außerdem wiederholen wir alle Tests mehrfach, um die Streuung des Ergebnisses in unsere Untersuchung einbeziehen zu können. Wie wir argumentieren werden, liegt in dieser Streuung der größte fundamentale (weil LLM-spezifische) Einwand gegen die Verwendbarkeit des Tools im Schulalltag.

Dies ist sodann aber nicht nur ein methodologischer Punkt unserer Untersuchung, sondern eine der wichtigsten Schlussfolgerungen auch für den Gebrauch des Korrekturtools: Erst wenn man ein und die selbe Schülerabgabe mehrfach korrigieren lässt, sieht man, wie sehr das Tool in seinem Output – also dem Feedback und der vorgeschlagenen Benotung – schwankt. Während Fobizz keiner Stelle transparent macht, dass die Ausgabe des Korrekturtools (von der für Schüler:innen mitunter viel abhängt) auf Zufallskomponenten beruht und deshalb zwischen Durchläufen variiert, ist es für Lehrkräfte höchstes Gebot, dieses Verhalten zu kennen und im Gebrauch einzukalkulieren.

**Vorgehensweise:** Zur Untersuchung des Korrekturtools haben wir eine exemplarische Aufgabenstellung zu einem Thema im Bereich Politik/Gesellschaftskunde entworfen. Sie lautete:

> Schreibe eine begründete Stellungnahme für oder gegen die Absenkung des Wahlalters auf 14 Jahre. Gehe dabei auf mindestens ein Argument für jede Seite ein und beziehe anschließend Position. (Schreibe ca. 150 - 250 Wörter.)

Den Erwartungshorizont haben wir durch die in Abbildung #fig:T:scope gezeigte Eingabe für die Box “Erwartungshorizont oder Musterlösung” des Korrekturtools konfiguriert. Die eingestellten Bewertungskriterien sind in Abbildung #fig:T:kriterien ersichtlich.

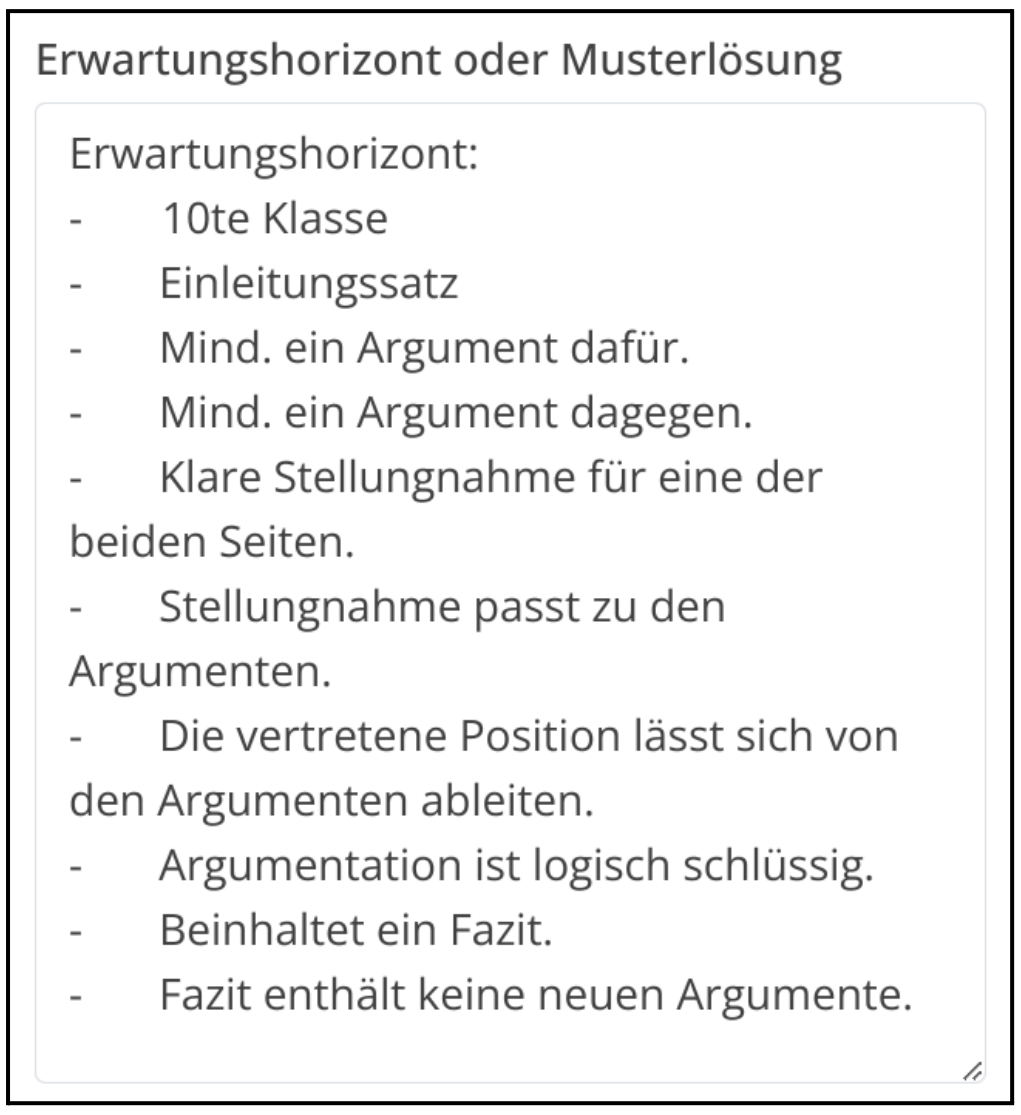

**Abb. #fig:T:scope:** Die Eingabe zum Erwartungshorizont für unsere Testreihen. Screenshot fobizz.com [28.11.2024].

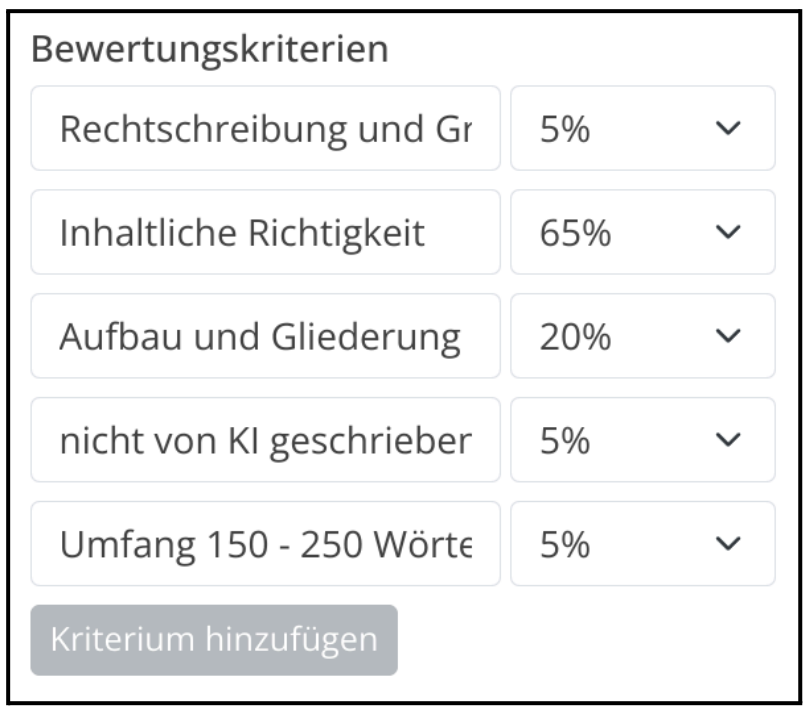

**Abb. #fig:T:kriterien:** Die Eingabe zu den Bewertungskriterien für unsere Testreihen. Screenshot fobizz.com [28.11.2024].

Als Bewertungsschema haben wir für unsere Tests aus den verfügbaren Maßstäben die “Skala von 1 bis 100%” gewählt (siehe Abbildung #fig:T:gradingscales). Für die Auswertung haben wir die Prozentbewertung teilweise nachträglich in Oberstufenpunkte umgerechnet und sind dann nach dem in Tabelle #tab:M:umrechnung angegebenen Schlüssel vorgegangen.

| klassische Schulnote | 1 | | | 2 | | | 3 | | | 4 | | | 5 | | | 6 |
|---|---|---|---|---|---|---|---|---|---|---|---|---|---|---|---|---|
| Punkte | 15 | 14 | 13 | 12 | 11 | 10 | 9 | 8 | 7 | 6 | 5 | 4 | 3 | 2 | 1 | 0 |
| Prozentschwelle (untere Grenze) | 95 | 90 | 85 | 80 | 75 | 70 | 65 | 60 | 55 | 50 | 45 | 40 | 33 | 27 | 20 | 0 |

**Tab. #tab:M:umrechnung:** Umrechnung von Prozentbewertungen in das Punktesystem der gymnasialen Oberstufe und in klassische Schulnoten.[38]

Das wie oben dargestellt für unsere Aufgabenstellung konfigurierte Korrekturtool haben wir im August 2024 zwei Testreihen unterzogen.

## Testreihe A

Zu unserer Aufgabenstellung haben wir 10 exemplarische Schülerabgaben erstellt, die ein Spektrum von sehr guten bis sehr schlechten Arbeiten abdecken (siehe Tabelle #tab:T:A; alle Texte sind im Materialanhang einsehbar). Wir haben jede dieser Abgaben in fünf unabhängigen Durchläufen durch das Korrekturtool bewerten lassen und die Resultate sowohl für dieselbe Abgabe als auch zwischen den Abgaben vergleichen. Im Augenmerk standen dabei – wie oben erläutert – Inkonsistenzen, interne Widersprüche und weitere von didaktischen Erwägungen weitestgehend unabhängige und in diesem Sinne objektive Fehler und Unzulänglichkeiten.

| Abgabe Nr. | Wörter | Qualität / Merkmale |
|---|---|---|
| 1 | 250 | gut geschrieben |
| 2 | 114 | zu kurz, nur Gegenargumente, nicht gut geschrieben, nur ein Absatz |
| 3 | 180 | schlecht geschrieben, Grenze zu Nonsense |
| 4 | 280 | zu lang, gut geschrieben |
| 5 | 228 | befriedigende Textqualität mit inhaltlichem Fehler (Wahlalter Europawahl ist nicht 14 Jahre) und unterschiedlichen eigenen Standpunkten am Anfang und am Ende |
| 6 | 256 | Etwas zu lang, kleine Fehler, ansonsten aber sehr gut |
| 7 | 41 | absolut low effort, grenzt an Arbeitsverweigerung, 3 Sätze |
| 8 | 69 | low effort, Thema verfehlt, vom Schwimmbadbesuch mit Oma erzählt |
| 9 | 170 | Mit ChatGPT generiert, kurz, prägnant. |
| 10 | 154 | gut bis befriedigend, Mittelmaß |

**Tab. #tab:T:A:** Liste der 10 in Testreihe A verwendeten Abgaben (simulierte Texte von Schüler:innen) mit Wortzahl und Angaben zur Qualität.

## Testreihe B

Mit 2 Abgaben (Nr. 1 und Nr. 10) haben wir in Testreihe B eine iterative Reihenuntersuchung angefertigt: Nach jedem Korrekturdurchlauf haben wir das Feedback des Korrekturtools zur Verbesserung der Abgabe eingearbeitet und den so verbesserten Text wieder korrigieren lassen. Dies erfolgte über 7–12 Iterationsstufen, um dabei zu beobachten, wie sich die numerische Bewertung und das inhaltliche Feedback durch die Verbesserungen verändern. Die Anzahl der durchgeführten Iterationsstufen hat sich daran orientiert, wie lange noch "gehaltvolles" Feedback kommt: Denn ab einem bestimmten Punkt beginnt sich das Feedback zu widersprechen oder nur noch "Fehler, die keine sind" zu enthalten; an diesem Punkt haben wir die Serie abgebrochen, indem wir, für eine letzte Verbesserung, den Text mit dem OpenAI KI-Tool ChatGPT final verbessert haben.

[38] Wir folgen hier https://de.wikipedia.org/wiki/Vorlage%3APunktesystem_der_gymnasialen_Oberstufe [abgerufen 2024-11-29].

# III. Ergebnisse

Wie in Abschnitt II erläutert, beschränken wir uns in dieser Untersuchung bewusst auf die Beobachtung solcher Unregelmäßigkeiten, die nicht von einer bestimmten didaktischen Auffassung in Bezug auf eine gute Bewertungspraxis abhängig sind. Das heißt, wir stellen lediglich objektive Funktionsdefizite des Korrekturtools heraus, die bereits *vor* einer didaktischen Beurteilung stehen und übergreifend anzuerkennen sind.

## Testreihe A

### 1. Zufälligkeit von Bewertungen und Rückmeldungen

#### a. Zufälligkeit der vorgeschlagenen Gesamtnote

Testreihe A umfasste die fünfmal wiederholte Erstellung einer Bewertung für jede der 10 exemplarischen Abgaben zu unserer Aufgabenstellung. Diese Methodologie gestattet es, das Tool hinsichtlich Konsistenz und Robustheit von Feedback und Bewertungen hin zu untersuchen. Abbildung #fig:E:volatilität visualisiert das für jede der Abgaben (X-Achse) resultierende Spektrum der empfohlenen Gesamtbewertung.

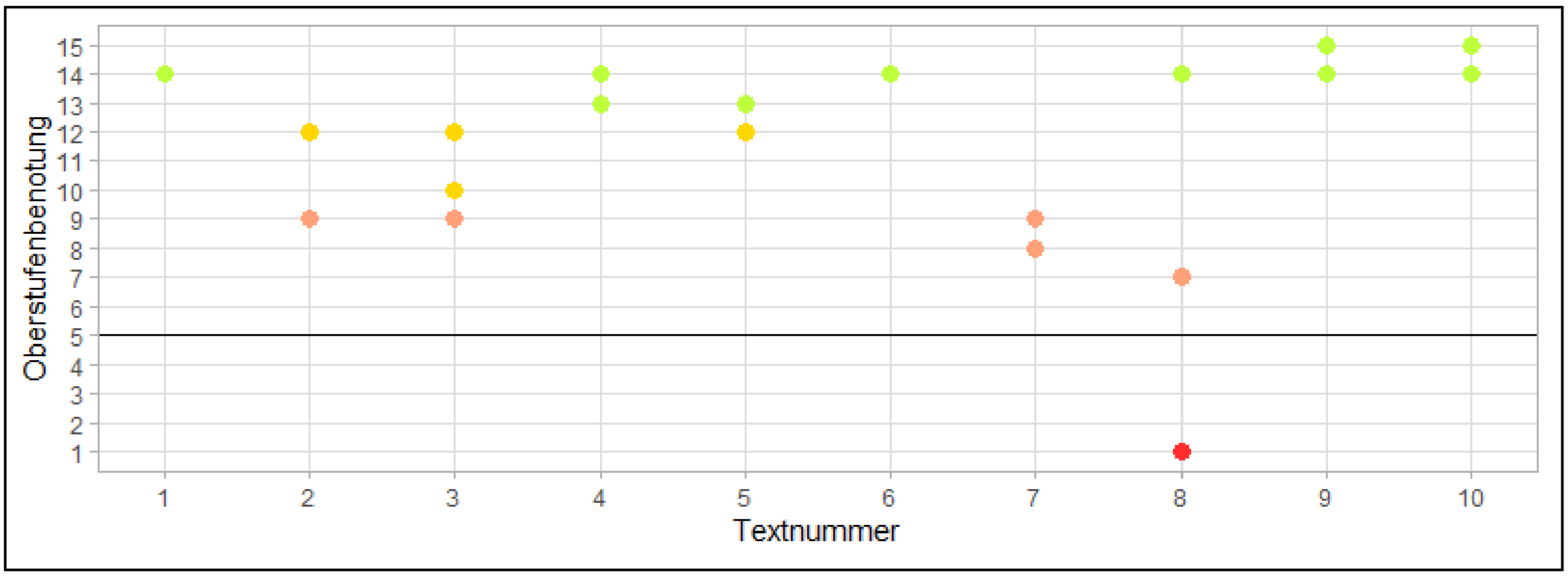

**Abb. #fig:E:volatilität:** Für jede simulierte Abgabe (X-Achse, nach Ordnungsnummer sortiert) werden vertikal untereinander die Gesamtnoten markiert (in Oberstufenpunkten, Y-Achste), die im Rahmen der fünf Korrekturdurchläufe jeweiligen Abgabe aufgetreten sind. Die Farbcodierung gibt die zugehörige klassische Schulnote an: grün = 1, gelb = 2, orange = 3, rot = 4 oder schlechter.[39]

[39] Wir haben die Anzahl Vorkommnisse derselben Bewertung innerhalb der fünf Korrekturdurchläufe pro Text bewusst nicht im Diagramm dargestellt (etwa durch größere Punkte oder mehrere Punkte nebeneinander). Damit möchten wir den sonst leicht entstehenden Eindruck vermeiden, es handle sich bei der auf der Y-Achse aufgetragenen Größe um eine statistische Größe, bei deren mehrfacher "Messung" man einzelne Abweichler ("Messfehler") streichen oder Erwartungswerte bilden könnte. Da im normalen Gebrauch des Tools eine mehrfache Bewertung desselben Texts nicht vorgesehen ist, liegt es nicht im normalen und intendierten Gebrauch des Tools, die vorgeschlagene Gesamtnote durch eine solche statistische Prüfung zu filtern.

Zu erkennen ist, dass bei 3 von 10 Abgaben die vorgeschlagene Bewertung über mehrfach wiederholte Durchläufen für dieselbe Lösung um mehr als eine Schulnote schwankt. In einem Fall schwankt sie von 1 Punkt bis 14 Punkte. Generell ist die vorgeschlagene Gesamtbewertung bei 8 von 10 der Arbeiten nicht stabil hinsichtlich der Wiederholung des automatisierten Bewertungsvorgangs, mit einer durchschnittlichen Schwankung um mehr als einen Punkt auf der 15-Punkte-Skala.

### b. Zufälligkeit der inhaltlichen Bewertung und qualitativen Rückmeldung

Nicht nur die numerische Benotung hat sich im Vergleich der Korrekturdurchläufe als stark schwankend erwiesen; selbes ist auch für die qualitative Bewertung und Würdigung zu beobachten. Die durch das Korrekturtool erstellte Bewertung enthält immer einen Abschnitt "Gesamtbewertung", welcher das Feedback summarisch zusammenfasst. Abbildungen #fig:E:6-3 und #fig:E:6-4 zeigen exemplarisch diese Absätze für Abgabe Nr. 6, Korrekturdurchläufe 6.3 und 6.4.

> Gesamtbewertung: 94%
>
> Der Text ist insgesamt sehr gut gelungen. Die Argumentation ist schlüssig und gut strukturiert. Es gibt nur wenige kleinere Fehler in der Rechtschreibung und Grammatik, die leicht zu korrigieren sind. Die inhaltliche Richtigkeit und der Umfang des Textes sind hervorragend.

**Abb. #fig:E:6-3:** Ausschnitt aus der Bewertung, Durchlauf 6.3 (Screenshot).

> Gesamtbewertung
>
> Der Text ist insgesamt gut gelungen und erfüllt die meisten Anforderungen. Es gibt einige kleinere Fehler in der Rechtschreibung und Grammatik, die leicht zu beheben sind. Die Argumentation ist logisch und gut strukturiert, und der Text liegt im geforderten Wortumfang.

**Abb. #fig:E:6-4:** Ausschnitt aus der Bewertung, Durchlauf 6.4 (Screenshot).

Gemäß Abbildung #fig:E:volatilität ist Abgabe 6 eine von nur zwei der zehn Abgaben, bei denen alle Korrekturwiederholungen denselben Notenvorschlag (14 Punkte) ergeben haben. Dennoch ist der Tenor des Feedbacks – hier am Vergleich der Gesamtbewertungen gut erkennbar – deutlich abweichend. In 6.4 ist die Terminologie ambivalent ("die meisten Anforderungen", "einige kleinere Fehler"), während sie in 6.3 eindeutiger positiv ist ("sehr gut gelungen", "schlüssig und gut strukturiert", "nur wenige kleinere Fehler", "hervorragend").

Es fällt außerdem auf, dass der Abschnitt "Gesamtbewertung" uneinheitlich auch eine Angabe der Gesamtnote enthält oder nicht enthält; siehe dazu weiter Ergebnis 4.c unten.

## 2. Unzuverlässige Erkennung inhaltlicher Defizite

### a. Falschbehauptungen

Unsere simulierten Schüler:innen-Abgaben enthielten teilweise simple inhaltliche Fehler. So hat beispielsweise Text 5 behauptet, das Wahlalter für die Europawahl sei kürzlich auf 14 Jahre abgesenkt worden (Abbildung #fig:E:5-2).

> Ich finde dass das Wahlalter auf 14 Jahren abgesenkt werden sollte. Hier schreibe ich warum ich das finde und warum manche Leute dagegen sind. Das Wahlalter ist ein umstrittenes Thema und auch in unserer Klasse haben wir viel diskutiert, vor allem weil letztens erst das Mindestalter für die Europawahl auf 14 Jahren abgesenkt wurde.

**Abb. #fig:E:5-2:** Screenshot Abgabe 5.

In keinem der fünf Korrekturdurchläufe für diese Abgabe wurde die faktisch falsche Behauptung des letzten Satzes als Fehler erkannt. Durchschnittlich erhielt der Text eine 88%-Bewertung in der Kategorie “inhaltlichen Richtigkeit”. Im Korrekturdurchlauf 5.3 wurde die inhaltliche Richtigkeit beispielsweise mit 90% bewertet und mit dem Feedback “Die Argumente sind inhaltlich korrekt und gut nachvollziehbar” versehen (Abbildung #fig:E:5-3).

| | | |
|---|---|---|
| Inhaltliche Richtigkeit | 90% | Die Argumente sind inhaltlich korrekt und gut nachvollziehbar. |

**Abb. #fig:E:5-3:** Ausschnitt aus der Bewertung, Durchlauf 5.3 (Screenshot).

### b. Nonsense, Arbeitsverweigerung, Thema verfehlt

Drei unserer Simulierten Abgaben stellen Arbeitsverweigerungen oder Verfehlungen des Themas dar: **Abgabe 3** (Gesamtbewertung zwischen 9 und 12 Punkte) begründet eine der Positionen mit der Meinung der Oma. **Abgabe 7** (Gesamtbewertung zwischen 8 und 9 Punkte) umfasst nur rund ein Drittel der geforderten Mindestwortzahl und führt als Hauptargument gegen die Absenkung des Wahlalters aus, dass 14-Jährige noch nicht so gut Kreuze setzen können wie Erwachsene. **Abgabe 8** (Gesamtbewertung zwischen 1 und 14 Punkten) verfehlt das Thema, indem von einem Schwimmbadbesuch erzählt wird.

Während diese drei Abgaben zwar im Durchschnitt über die mehrfache Wiederholung des automatisierten Bewertungsvorgangs die schlechteste Bewertungen erhielten, liegen zwei dieser Abgaben (3 und 7) klar im Notenbereich gut/befriedigend und hätten damit in der automatisierten Bewertung mühelos bestanden – was angesichts der Qualität zweifelhaft erscheint. Die das Thema verfehlende Abgabe 8 erhält bei Wiederholung der automatisierten Bewertung das größte Notenspektrum (1 bis 14 Punkte). In Korrekturlauf 8.1 wird die Lösung mit 14 Punkten bewertet und erhält ein positives, wenn auch unspezifisches Feedback (Abbildung #fig:E:8-1).

> **Gesamtbewertung**
>
> Der Text ist insgesamt gut geschrieben und verständlich. Es gibt nur wenige grammatikalische Fehler, und der Inhalt ist korrekt und anschaulich beschrieben. Allerdings ist der Text zu kurz und erfüllt nicht die geforderte Wortanzahl. Achte darauf, die Sätze abwechslungsreicher zu gestalten und die korrekte Verwendung von Präpositionen und Konjunktionen zu beachten.

**Abb. #fig:E:8-1:** Ausschnitt aus der Bewertung, Durchlauf 8.1 (Screenshot).

## 3. Unzuverlässige Umsetzung einzelner Bewertungskriterien

Das Korrekturtool bietet eine freie Eingabemöglichkeit für Bewertungskriterien (siehe Abbildung #fig:T:kriterien) durch die Lehrkraft, inklusive relativer Gewichtung der Kriterien für die Gesamtnote. So wirbt Fobizz in einer Video-Anleitung zum Korrekturtool explizit mit der Möglichkeit, nach Belieben eigene Bewertungskategorien einzusetzen.[40] Die Umsetzung und Berücksichtigung dieser Kriterien bei der Bewertung ist allerdings unzuverlässig und intransparent, wie die folgenden Beobachtungen zeigen.

### a. Textlänge als Kriterium

Unsere exemplarische Aufgabenstellung enthält einen vorgegebenen Wortumfang von 150 bis 250 Wörtern und wir haben die Befolgung als ein Bewertungskriterium eingestellt (siehe Abbildung #fig:T:kriterien). Unter den zehn exemplarischen Abgaben gibt es fünf, deren Wortumfang außerhalb der vorgegebenen Grenzen liegt: 2, 7 und 8 sind zu kurz; 4 und 6 sind zu lang (siehe Tabelle #tab:T:A). Bei den Abgaben 2, 7 und 8, also bei den zu *kurzen* Texten, wurde dafür in sämtlichen Korrekturdurchläufen eine niedrigere Bewertung gegeben; die Bewertungen reichten hier von 30% bis

[40] https://tools.fobizz.com/ai/feedbacks/info [abgerufen 2024-11-29].

60% (entspricht 3–7 Punkte). Die Überlänge von 4 und 6 wurde in keinem der Korrekturdurchläufe erkannt und die Teilbewertung für das entsprechende Kriterium war stets 100%.

| | | |
|---|---|---|
| Umfang 150 - 250 Wörter | 60% | Der Text ist mit 123 Wörtern zu kurz und erfüllt nicht die geforderte Wortanzahl. |

**Abb. #fig:E:2-5:** Ausschnitt aus der Bewertung, Durchlauf 2.5 (Screenshot).

Da die Einhaltung des Umfangs zu den vorgegebenen Bewertungskriterien gehörte, wurde eine entsprechende Teilnote mit "Feedback" bei allen Korrekturläufen in der Tabelle "Bewertung" angeführt. In 14 von 50 Durchläufen (2.1–5, 6.3, 7.1–5, 9.1, 9.3–4) wurde die Textlänge im Feedback quantifiziert, siehe exemplarisch im Durchlauf 2.5 (Abbildung #fig:E:2-5), wo es heißt: "Der Text ist mit 123 Wörtern zu kurz". In nur 2 der 14 Fälle mit Quantifizierung ist die angegebene Zahl korrekt (nämlich bei 7.3–4). In 5 weiteren Fällen (Durchläufe 8.1–5) wird eine faktisch abweichende Textlänge als in der "Fehlerliste" als Fehler angeführt, jedoch ohne Angabe der Wortzahl (siehe Abbildung #fig:E:8-2).

| | | |
|---|---|---|
| Umfang 150 - 250 Wörter | 50% | Der Text ist zu kurz und erfüllt nicht die geforderte Wortanzahl. |

**Abb. #fig:E:8-2:** Ausschnitt aus der Bewertung, Durchlauf 8.2 (Screenshot).

Schließlich gibt es in einigen Durchläufen den Fall, dass die Textlänge negativ beanstandet wird, obwohl die Abgabe innerhalb des geforderten Umfangs liegt (Durchläufe 3.1–3, 3.5, 5.2). In diesen Fällen wird eine Bewertung von 90% gegeben, mit der Begründung "Der Text liegt im geforderten Umfang, könnte aber etwas ausführlicher sein." (exemplarisch Durchlauf 3.3, siehe Abbildung #fig:E:3-3).

| | | |
|---|---|---|
| Umfang 150 - 250 Wörter | 90% | Der Text liegt im geforderten Umfang, könnte aber etwas ausführlicher sein. |

**Abb. #fig:E:3-3:** Ausschnitt aus der Bewertung, Durchlauf 3.3 (Screenshot).

Insgesamt wird die Einhaltung eines bestimmten Textumfangs unzuverlässig bewertet. Insbesondere bei Texten, die den Umfang nicht einhalten, wird dies nur in etwa der Hälfte der Fälle erkannt. Siehe zur Übersicht das Diagramm in Abbildung #fig:E:umfang-diagramm.

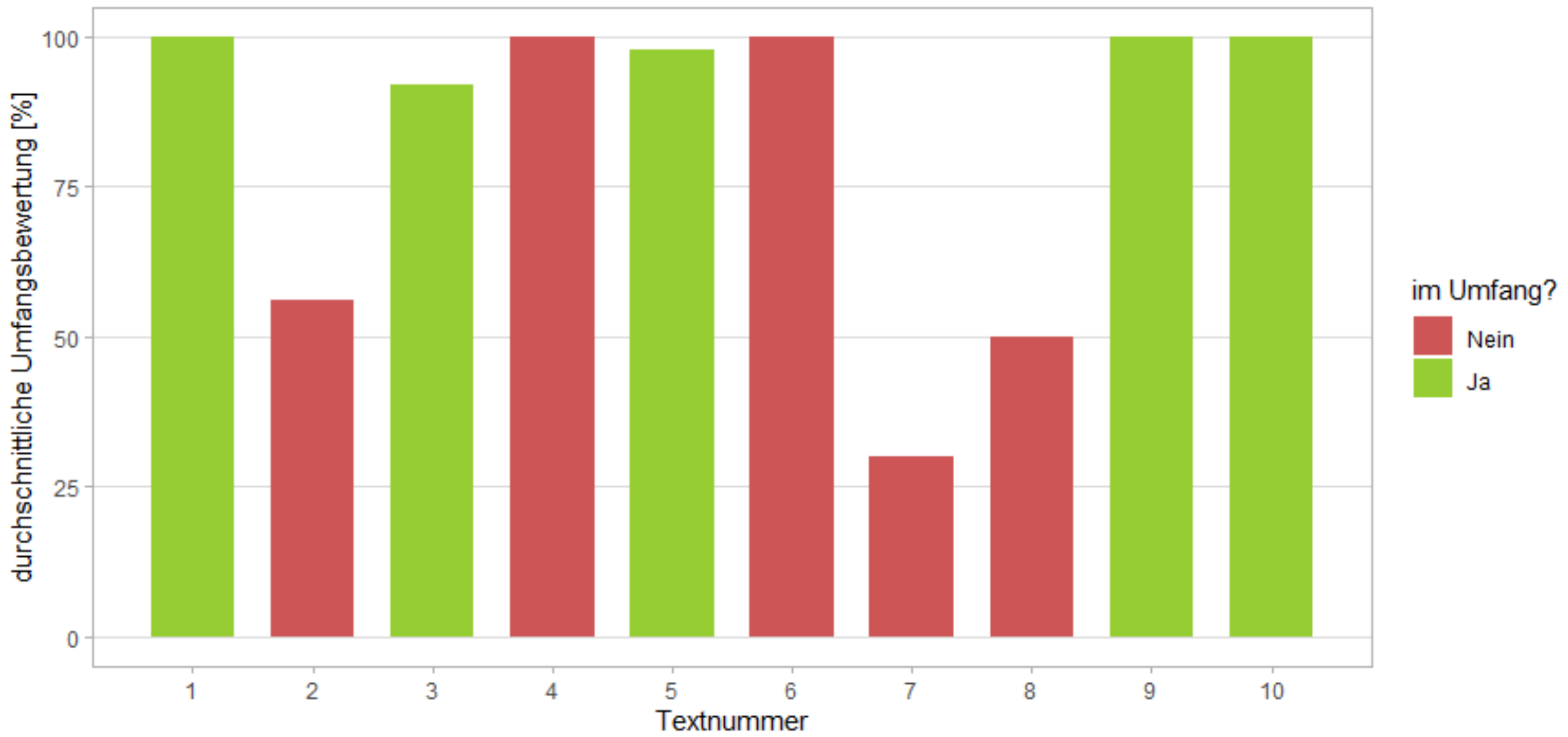

**Abb. #fig:E:umfang-diagramm:** Für die einzelnen Abgaben (X-Achse) die durchschnittliche Bewertung in der Kategorie "Umfang" über die fünf unabhängigen Korrekturdurchläufe. Farbcodierung: Grün = Abgabe liegt tatsächlich im geforderten Umfang; rot = Abgabe liegt außerhalb des geforderten Umfangs.

### b. KI-generierte Abgaben erkennen

Zu den von uns eingestellten Bewertungskriterien gehörte auch die Kategorie “Nicht von KI geschrieben”, um überprüfen zu lassen, ob der Text unter Zuhilfenahme von KI-Tools wie ChatGPT erstellt wurde. Bekanntermaßen ist die automatisierte Erkennung von KI-generierten Texten ein ungelöstes Problem.[41] Auch Fobizz erklärt in einem Blogpost auf seiner Webseite, dass so eine Erkennung nicht möglich ist.[42] Wenn jedoch eine Nutzer:in, wie wir es getan haben, “Nicht von KI geschrieben” in die freie Liste von Bewertungskriterien einfügt, dann erzeugt das Tool ohne entsprechenden Hinweis ein Feedback, in dem das Bewertungskriterien aufgeführt und mit einer Teilnote versehen wird. Das Feedback suggeriert damit, dass in dieser Kategorie wirklich eine Beurteilung vorgenommen wurde. In unseren Tests liegt die Bewertung in dieser Kategorie in allen 50 Durchläufen bei 100%, versehen mit einem Kommentar, dass der Text authentisch und nicht von einer KI generiert worden sei (siehe exemplarisch Abbildung #fig:E:9-3 für den Durchlauf 9.3). Dies wird auch bei Text 9 behauptet, der tatsächlich mithilfe von ChatGPT erstellt wurde.

| | | |
|---|---|---|
| Nicht von KI geschrieben | 100% | Der Text wirkt authentisch und nicht von einer KI generiert. |

**Abb. #fig:E:9-3:** Ausschnitt aus der Bewertung, Durchlauf 9.3 (Screenshot).

## 4. Inkonsistentes Feedback

### a. Uneinheitliche und zufällige Bezeichnung von Fehlerkategorien

Die Zuordnung der erkannten Fehler zu jeweils einer Fehlerkategorie (dritte Spalte der tabellarischen Auflistung “Fehlerliste”, siehe Abbildung #fig:E:bewertung) ist ein integraler Bestandteil des automatisch generierten Feedbacks. Dabei fällt im Vergleich verschiedener Korrekturdurchläufe nicht nur die uneinheitliche Benennung der dritten Spalte dieser tabellarischen Ausgabe auf (“Fehlerart”, “Fehlertyp”), sondern viel mehr noch eine stark variierende Benennung der Fehlerkategorien. Anhand der Sichtung von 50 Bewertungen haben wir die folgenden Benennungen von Fehlerkategorien extrahiert. Die unten vorgeschlagene Clusterung der Benennungen bildet Synonym-Cluster, aus denen das Korrekturtool bei jedem Durchlauf ohne erkennbares Prinzip, also scheinbar zufällig, einen Begriff auswählt.

- “Rechtschreibung”, “Rechtschreibung (Groß-/Kleinschreibung)”, “Groß-/Kleinschreibung”,
- “Abkürzung”, “Typografie”, “Typografie (Doppeltes Leerzeichen)”
- “Grammatik”, “Grammatikfehler”, “Konjunktiv”
- “Sprachstil”, “Stilistischer Fehler”, “Stil”, “Wiederholung”, “Satzbau”, “Ausdruck”, “Wortwahl”
- “Unklarer Bezug”, “Bezug”
- “Argumentationsfehler”, “Unpräzise Argumentation”
- “Unpräzise Formulierung”, “Klarheit”
- “Keine Fehler gefunden” (siehe auch 4.b)

Die volatile Kategorisierung der Fehler bzw. Benennung der Fehlertypen verunmöglicht die Vergleichbarkeit zwischen einzelnen Korrekturen und schafft Unklarheit über die zugrundeliegenden Bewertungskriterien.

### b. Fehler, die keine sind

Die automatisch erstellte Bewertung des Korrekturtools enthält stets eine tabellarische Auflistung von Fehlern, bestehend aus den Spalten “Fehler”, “Korrektur” und “Fehlerart” (wobei die letzte Spalte nach keinem erkennbaren Prinzip teilweise auch mit “Fehlertyp” überschrieben ist). In 10 der 50 getesteten

[41] https://link.springer.com/article/10.1007/s40979-023-00146-z ; https://arxiv.org/abs/2303.11156 ; https://link.springer.com/article/10.1007/s40979-023-00140-5
[42] https://fobizz.com/warum_ki-detektoren_nichts_bringen/ [abgerufen 2024-11-29].

Durchläufe (4.1–5, 9.1–5) kommt es vor, dass darin Fehler aufgeführt werden, die keine sind (siehe exemplarisch Abbildungen #fig:E:9-1 und #fig:E:4-1).

Fehlerliste

| Fehler | Korrektur | Fehlerart |
|---|---|---|
| "Ein Argument dafür ist, dass Jugendliche heute durch das Internet und soziale Medien gut informiert und politisch engagiert sind." | "Ein Argument dafür ist, dass Jugendliche heute durch das Internet und soziale Medien gut informiert und politisch engagiert sind." | Keine Fehler gefunden |

**Abb. #fig:E:9-1:** Ausschnitt aus der Bewertung, Durchlauf 9.1 (Screenshot).

| | | |
|---|---|---|
| "16 Jahre gesenkt" | "16 Jahre gesenkt" | Typografie |

**Abb. #fig:E:4-1:** Ausschnitt aus der Bewertung, Durchlauf 4.1 (Screenshot).

Dieses Phänomen tritt in zwei Varianten auf: als Fehlerkategorie kann eine (nicht zutreffende oder nicht nachvollziehbare) existierende Fehlerkategorie angegeben werden (in Abbildung #fig:E:4-1: “Typographie”) oder die Irrkategorie “Kein Fehler gefunden” (Abbildung #fig:E:9-1).

#### c. Widersprüchliche Angabe der Gesamtnote

Wie beschrieben erwähnt, enthält die automatisierte Bewertung immer einen Abschnitt “Gesamtbewertung”, welcher unregelmäßig in einigen Durchläufen die vorgeschlagene Gesamtnote mit anführt und in anderen nicht (vgl. Abbildungen #fig:E:1-3 und #fig:E:1-4). Dieses inkonsistente Verhalten tritt noch in einer verschärften Form auf, wenn die vorgeschlagene Gesamtnote *mehrfach* im Bewertungsdokument angeführt wird (dies ist der Fall bei den Durchläufen 3.2, 3.3 und 6.3). In diesen Fällen kommt es mehrheitlich vor, dass die angegebene Gesamtnote innerhalb des Dokuments widersprüchlich ist, die Note also nicht identisch wiederholt wird (dies ist der Fall bei 3.2 und 3.3, siehe exemplarisch Abbildung #fig:E:3-3).

Gesamtbewertung: 83%

Der Text zeigt eine gute Basis, aber es gibt Raum für Verbesserungen in der Argumentation und Grammatik. Achte darauf, deine Argumente klarer und logischer zu formulieren und weniger auf persönliche Meinungen zu stützen.

Gesamtnote

74.00

**Abb. #fig:E:3-3:** Ausschnitt aus der Bewertung, Durchlauf 3.3 (Screenshot).

## Testreihe B

### 5. Die Umsetzung des Feedbacks führt nicht zur Verbesserung

#### a. Fehlende Monotonie bei der Einarbeitung von Vorschlägen

In Testreihe B haben die Abgaben Nr. 1 und 10 einer Serie von Verbesserungen und automatisierten Bewertungen unterzogen: Beginnend bei der Originalabgabe haben wir stets die Verbesserungsvorschläge aus der Fehlerliste der automatisierten Bewertung umgesetzt und den Text dann neu bewerten lassen. Dies erfolgte über 7–12 Iterationsstufen, wobei wir für die letzte Iterationsstufe den Text mittels des frei zugänglichen und bei Schüler:innen sehr beliebten KI-Tools “ChatGPT” von OpenAI haben verbessern lassen (was einer Täuschung gleichkommt).

Abbildung #fig:E:irrfahrt zeigt den Verlauf der Gesamtbewertung der beiden Texte in der jeweiligen Iterationsstufe der sukzessiven Verbesserungen (die einzelnen Textversionen und automatisierten Bewertungen können im Materialanhang eingesehen werden). Es fällt auf, dass die Gesamtbewertung in beiden Fällen nach den ersten Verbesserungsläufen zunächst rückläufig ist. Erst in der vierten Iteration übertreffen beide Texte leicht die numerische Ausgangsbewertung des Tools. Auch dieser Zustand ist nicht stabil; bei weiterer Umsetzung der Verbesserungsvorschläge geht in beiden Fällen die Gesamtnote wieder zurück und fällt deutlich unter das Ausgangsniveau. Berücksichtigt man die letzte Iteration (mit ChatGPT überarbeitete Texte) nicht, so liegt der Durchschnitt der Gesamtbewertungen aller Iterationsstufen für Text 1 bei 93,96% und für Text 10 bei 91,15%. Dies ist in beiden Fällen unterhalb des Ausgangsniveaus (94,0% für Text 1 und 92,25% für Text 10).

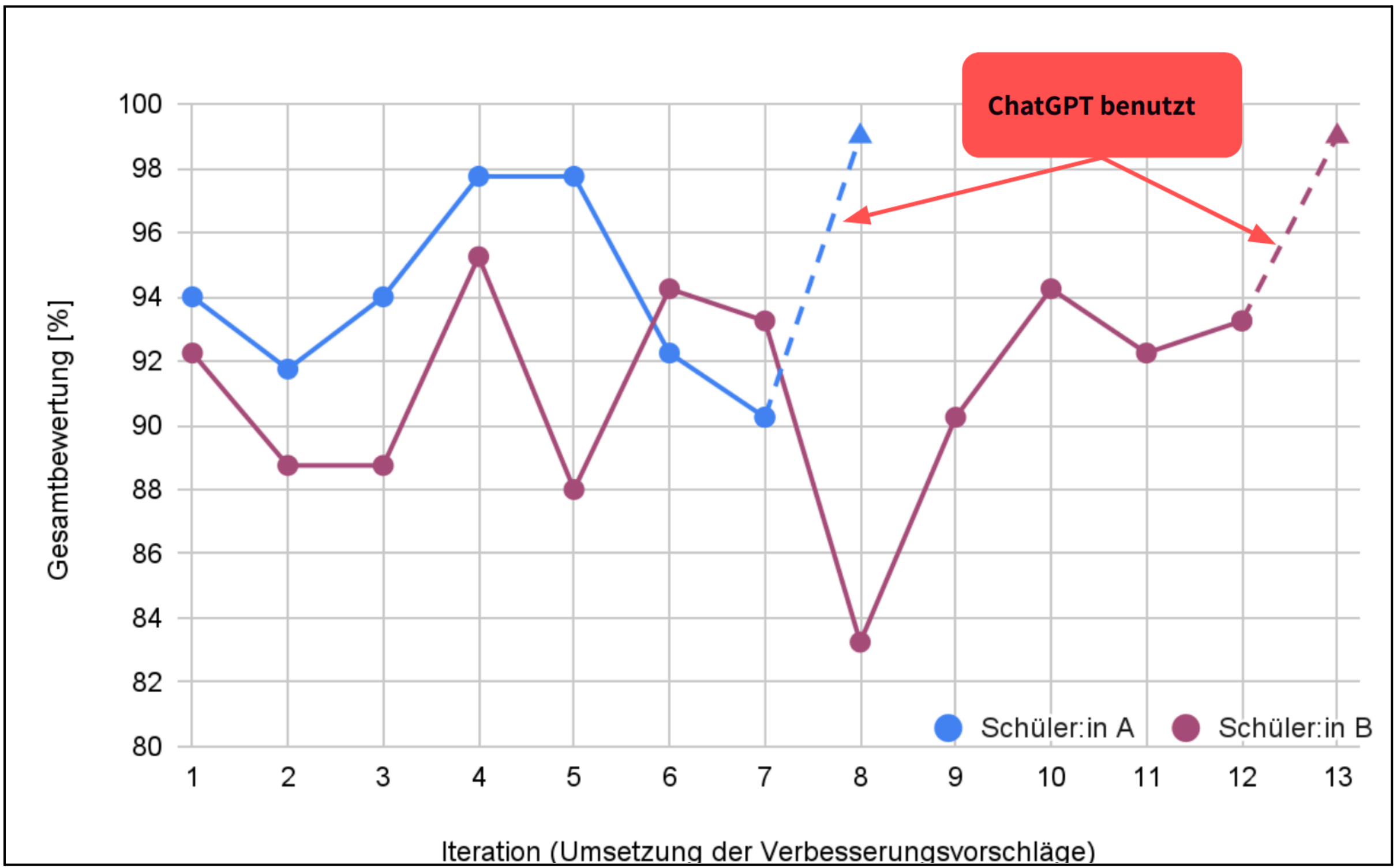

**Abb. #fig:E:irrfahrt:** Gesamtbewertung bei iterativer Umsetzung der Verbesserungsvorschläge des Korrekturtools für die Abgaben Nr. 1 (Schüler:in A) und 10 (Schüler:in B). Die mit einem Dreieck ▲ gekennzeichneten Werte (Abschluss der iterativen Reihe) entsprechen der Bewertung der Texte, die final mithilfe von ChatGPT überarbeitet wurden.

### b. Irrfahrt-Charakter des qualitativen Feedbacks

Bereits dem Verlauf der numerischen Gesamtbewertung in Abbildung #fig:E:irrfahrt ist ein “flatterhafter” Charakter zu eigen – wir bezeichnen das als “Irrfahrt”. Selbige Beobachtung gilt auch für das qualitative Feedback in Form der Fehlerliste und der Abschnitte “Was du gut gemacht hast” bzw. “Was du verbessern kannst”. So ist im Verlauf der sukzessiven Umsetzung der Verbesserungsvorschläge stets ein Punkt zu erkennen, ab dem die Rückmeldungen “im Kreis fahren”.

| | | |
|---|---|---|
| "Dies zeigt sich beispielsweise darin, dass Jugendliche tendenziell häufiger extreme Parteien wählen." | "Dies zeigt sich beispielsweise darin, dass Jugendliche tendenziell häufiger extreme Parteien wählen könnten." | Spekulation/Verallgemeinerung |

**Abb. #fig:E:irrfart-10-5:** Ausschnitt aus der “Fehlerliste”, Testreihe B, Text 10, Iteration 5 (Screenshot).

| | | |
|---|---|---|
| "Dies zeigt sich beispielsweise darin, dass Jugendliche tendenziell häufiger extreme Parteien wählen könnten." | "Dies zeigt sich beispielsweise darin, dass Jugendliche tendenziell häufiger extreme Parteien wählen." | Konjunktiv statt Indikativ |

**Abb. #fig:E:irrfart-10-6:** Ausschnitt aus der "Fehlerliste", Testreihe B, Text 10, Iteration 6 (Screenshot).

| | | |
|---|---|---|
| "Dies zeigt sich beispielsweise darin, dass Jugendliche tendenziell häufiger extreme Parteien wählen." | "Dies zeigt sich beispielsweise darin, dass Jugendliche tendenziell häufiger extreme Parteien wählen könnten." | Konjunktiv zur Verdeutlichung der Möglichkeit |

**Abb. #fig:E:irrfart-10-7:** Ausschnitt aus der "Fehlerliste", Testreihe B, Text 10, Iteration 7 (Screenshot).

Abbildungen #fig:E:irrfart-10-5, #fig:E:irrfart-10-6 und #fig:E:irrfart-10-7 zeigen diese Beobachtung exemplarisch anhand eines Ausschnitts der Fehlerliste für die Iterationen 5–7 zu Text 10, in denen ein Formulierungsdetail inkonsistent hin und her geändert wird.

### 6. Erreichbarkeit einer Bestbewertung nur durch Einsatz von ChatGPT (Täuschung)

In beiden Serien haben wir nach 7–12 Iterationen einen Punkt erreicht, an dem das weitere Einarbeiten der Rückmeldungen nicht sinnvoll zu sein schien, weil die Rückmeldungen zwischen zwei Optionen oszillierten (siehe vorherigen Punkt) und sich in marginalen Details festgebissen hatten. An diesem Punkt wurde für eine finale Überarbeitung ChatGPT verwendet, Abbildung #fig:E:chatgpt-prompt zeigt exemplarisch den dafür in Fall von Text 1 verwendeten Prompt.

> Ich muss für den Politikunterricht in der 10ten Klasse folgende Aufgabe bearbeiten: "Schreibe eine begründete Stellungnahme für oder gegen die Absenkung des Wahlalters auf 14 Jahre. Gehe dabei auf mindestens ein Argument für jede Seite ein und beziehe anschließend Position. (Schreibe ca. 150 - 250 Wörter.)"
> Kannst du diesen Text den ich bereits geschrieben habe bewerten und anschließend verbessern?
> *[Text eingefügt.]*

**Abb. #fig:E:chatgpt-prompt:** ChatGPT Prompt zur Verbesserung von Text 1 für die Iteration 8.

Zu beobachten ist in beiden Fällen, wie erst bei der Verwendung von ChatGPT eine nahezu perfekte numerische Bewertung erteilt wird. Zu bedenken ist, dass das ChatGPT zugrundeliegende große Sprachmodell GPT-4 auch dem Fobizz Korrekturtool zugrunde liegt.

## IV. Diskussion

Im Folgenden diskutieren wir die oben relativ technisch zusammengetragenen Beobachtungen hinsichtlich ihrer Auswirkungen auf den Bildungsalltag und ihrer Bedeutung für die Qualität und Benutzbarkeit des Korrekturtools. Dies gilt zugleich als eine übersichtliche Zusammenfassung der beobachteten Mängel. In einem folgenden Abschnitt werden wir herausstellen, dass nur die wenigsten der Mängel eine technische Lösung erwarten lassen, während die meisten auf prinzipielle Limitationen von großen Sprachmodellen (LLMs) zurückgehen.

## Bedeutung und Bewertung der beobachteten Mängel

### 1. Zufälligkeit von Bewertungen und Rückmeldungen

Indem wir jede Abgabe für unsere exemplarische Aufgabenstellung fünfmal durch das Korrekturtool bewerten lassen haben, konnten wir beobachten, dass sowohl der Notenvorschlag als auch die qualitative inhaltliche Rückmeldung zwischen den verschiedenen Bewertungsdurchläufen *für ein und dieselbe Lösung* teilweise erheblichen Schwankungen unterlagen. Nur bei 2 von 10 Abgaben war die empfohlene Gesamtnote über die fünf Wiederholungen stabil; bei 3 von 10 schwankte die Bewertung über mehr als eine Schulnote, in einem Fall sogar zwischen 1 und 14 Punkten. Auch das qualitative Feedback unterlag erheblichen Schwankungen, dies auch bei Bewertungsdurchläufen, die auf dieselbe numerische Bewertung hinausliefen – was bedeutet, dass die Stabilität der Note kein Garant für die Stabilität des inhaltlichen Feedbacks ist.

Diese Beobachtungen stellen einen gravierenden objektiven Qualitätsmangel des Korrekturtools dar, der durch Lehrkräfte, die es verwenden, zwingend berücksichtigt werden muss: **Wer diese Eigenschaft des Tools nicht kennt, vergibt mehr oder weniger ausgewürfelte Noten und mehr oder weniger ausgewürfelte Rückmeldungen.** Die Tücke an diesem Mangel ist, dass die Benutzer:in ihn erst erkennt, wenn sie auf die (wohl unwahrscheinliche und überdies zeitraubende) Idee kommt, dieselbe Abgabe mehrfach bewerten zu lassen und die Resultate zu vergleichen.

Die Benutzerführung des Tools enthält keinen Hinweis auf die zufallsabhängige Volatilität von Feedback und Bewertung. Im Gegenteil: Die Vermarktung und Beschreibung des Korrekturtools suggeriert, dass das Tool kriteriengeleitete Rückmeldungen und Benotungen erstellt. Der Einsatz von KI zur Bewertung wird sowohl im allgemeinen Diskurs über KI im Schulunterricht als auch konkret von Fobizz sogar als *Garant für mehr Objektivität* in der Bewertung der Schüler:innen beschrieben, da durch die Automatisierung vermeintlich persönliche Vorurteile der Lehrkräfte ausgeschaltet werden würden. Biases und Vorurteile der Lehrkraft fallen in diesem Vergleich allerdings gar nicht ins Gewicht, wenn die automatisierte Bewertung aus rein statistischen Gründen um mehrere Notenpunkte schwanken kann, was überhaupt nur erkenntlich wird, wenn die Lehrer:in auf die Idee kommt, mehrere Bewertungsdurchläufe zu unternehmen.

Zwar erhalten die Bewertungsdokumente, die vom Korrekturtool erstellt werden, einen Warnhinweis:

> Gut zu wissen:
>
> Die Antworten der KI sind als **Vorschläge** zu verstehen. Sie können niemals eine fundierte didaktische Beurteilung der Schülerleistungen ersetzen. (Übersetzung R.M.; siehe Abbildung #fig:D:disclaimer)

**Good to know:**

The AI's answers are to be understood as **suggestions**. They can never replace a well-founded didactic assessment of student performance.

**Abb. #fig:D:disclaimer:** Screenshot Bewertungsdokument (PDF-Output).

Dieser Warnhinweis klärt jedoch nicht über die hier diagnostizierte Eigenschaft auf, dass die qualitativen wie numerischen Ausgaben des Tools *statistischen Fluktuationen* unterliegen, die erst bei wiederholter Verwendung für ein und dieselbe Abgabe erkennbar sind. Das hat mit "fundierter didaktischer Beurteilung" wenig zu tun: Es handelt sich bei diesen Fluktuationen um typische und prinzipielle Eigenschaften von Systemen wie ChatGPT und den zugrundeliegenden großen Sprachmodellen (LLMs), in denen bei der Generierung der "Antworten" auf einen Prompt algorithmische Zufallsentscheidungen im Spiel sind.

Zum Vergleich: Die Benutzerführung des bekannten Chatbot-Systems ChatGPT (OpenAI) enthält gezielt einen Button zum “neu generieren” der Antwort auf einen Prompt. Über diese Funktion kann die Benutzer:in bei einer nicht zufriedenstellenden Antwort nochmal eine Variation erzeugen lassen.

Es scheint dem Wertversprechen automatisierter Erzeugung von Rückmeldungen und Bewertungen von Schularbeiten jedoch entgegen zu laufen, dass die Lehrkraft so lange die Bewertung neu generieren lassen kann – und bei verantwortungsvoller Verwendung sogar *muss* –, bis das Resultat möglicherweise ihrem didaktisch fundierten Urteil entspricht.

## 2. Unzuverlässige Erkennung inhaltlicher Defizite

Wir haben systematisch beobachtet, dass das Korrekturtool faktische Falschbehauptungen (z.B.: “Das Wahlalter in der EU ist auf 14 Jahre gesenkt worden.”) und Fälle offensichtlicher Arbeitsverweigerung (Nonsense-Abgaben) nicht verlässlich erkennt. Weniger die sehr guten Abgaben, als gerade “low effort” und absichtlich “unsinnige” Abgaben, die am ehesten einem Troll- oder Verweigerungsverhalten entsprechen und deren angemessene Bewertung notorisch schwierig ist, führen zu starken Schwankungen der numerischen Bewertung (teils über mehrere Schulnoten) und des qualitativen Feedbacks zwischen den Bewertungsdurchläufen.

Das Bewerten “schlechter” Lösungen ist schon immer schwieriger und nervenzehrender als das Bewerten guter Lösungen. Hier scheint die Automatisierung des Korrekturtools keine Ausnahme zu bilden und genaues Überprüfen der Bewertungsvorschläge ist hier nötig. Während das Erkennen oder Nichterkennen eindeutiger Falschbehauptungen (Mangel 2.a) trivialerweise einen Mangel darstellt, ist die Unzuverlässigkeit in der *qualitativen* Rückmeldung auf nicht falsche, aber unbemühte, “wirre” oder arbeitsverweigernde Abgaben (Magel 2.b) das komplexere Problem.

Lehrer:innen können sich gerade in diesen, didaktisch und pädagogisch oft anspruchsvollen Fällen nicht auf das Korrekturtool verlassen. Werden Vorschläge für Rückmeldungen oder Bewertungen unhinterfragt übernommen, besteht ein großes Risiko der Delegitimierung des Prinzips von Bewertung und Benotung: denn bei zufallsbedingt zu guter Bewertung wird Arbeitsverweigerung nicht erkannt; bei zufallsbedingt zu schlechter Bewertung werden vorhandene Bemühungen nicht honoriert.

## 3. Unzuverlässige Umsetzung einzelner Bewertungskriterien

Das Korrekturtool erlaubt die freie Eingabe einer beliebigen Liste von stichwortartig bezeichneten Bewertungskriterien zusammen mit einer relativen prozentualen Gewichtung der Kriterien für die Gesamtnote (siehe Abbildung #fig:T:kriterien). Dieses Feature wird in den begleitenden Video-Tutorials von Fobizz explizit beworben.[43] Unsere Untersuchen haben gezeigt, dass nicht jedes Kriterium, das man in diese Liste eintragen könnte, gleichermaßen zuverlässig von der Maschine beherrscht wird. So konnten wir zeigen, dass etwa ein Kriterium, das sich auf die Einhaltung von Vorgaben über den Textumfang bezieht, in weniger als 50% der Fälle korrekt umgesetzt wurde (Mangel 3.a). Das Kriterium, KI-generierten Text negativ zu bewerten, wurde überhaupt nicht korrekt umgesetzt (Mangel 3.b).

Es ist über große Sprachmodelle wie GPT-4, das dem Korrekturtool zugrunde liegt, wissenschaftlich bekannt, dass sie nicht gut Wörter “zählen” und kaum KI-generierten Text erkennen können. Dass diese Bewertungskriterien also nicht korrekt umgesetzt werden, wundert nicht. Entscheidend ist hier jedoch der Umgang damit: Weder weist das Benutzerinterface darauf hin, dass bestimmte Kriterien nicht sinnvoll in die Liste eingetragen werden können, weil das Tool sie nicht beherrscht, noch findet sich in der ausgegebenen Bewertung ein entsprechender Hinweis darauf, dass die verschiedenen berücksichtigten Kriterien von dem Tool unterschiedlich gut beherrscht (und teilweise gar nicht) beherrscht werden.

[43] https://tools.fobizz.com/ai/feedbacks/info

Benutzerführung, Beschreibung des Tools und die produzierten Ausgaben (Bewertungsdokumente) suggerieren zweifelsfrei, dass beliebige Kriterien eingegeben und dann auch umgesetzt werden können. Die Insuffizienz des Tools bei bestimmten Kriterien bleibt somit strategisch verborgen; es wird das Bild eines omnipotenten KI-Tools kreiert. Somit liegt hier ein Beispiel für Verantwortungslosigkeit im Interaktionsdesign vor. Es wäre angemessen und richtig, dass das Tool die Verarbeitung von Kriterien, die es nicht beherrscht, verweigert, oder transparent Auskunft darüber gibt, wie gut es die einzelnen angegebenen Kriterien beherrscht. Nutzer:innen, die die besonderen Eigenschaften von LLMs nicht kennen, fallen möglicherweise sehr leicht auf diese intransparente Kommunikation hinein.

## 4. Inkonsistentes Feedback

In der Zusammenschau von 50 Bewertungsdurchläufen für 10 Abgaben zu unserer exemplarischen Aufgabenstellung haben wir diverse Glitches in der Form von Inkonsistenzen und internen Selbstwidersprüchen der erstellten Bewertungsdokumente festgestellt. Die Fehlerkategorien, die in der Tabelle "Fehlerliste" verwendet werden, sind uneinheitlich (z.B. "Sprachstil", "Stilistischer Fehler", …, siehe Mangel 4.a). Es wurden in 10 der 50 Bewertungsdurchläufe Fehler aufgelistet, die keine sind – in diesen Fällen sind die beanstandeten und die korrigierten Textstellen, wie sie in der Fehlerliste angegeben werden, identisch (Mangel 4.b). In 2 der 50 Bewertungsdurchläufe wurden schließlich innerhalb des Bewertungsdokuments unterschiedliche Gesamtnoten angegeben (Mangel 4.c).

Die Uneinheitlichkeit der verwendeten Terminologie für Fehlertypen erzeugt bei Lehrenden und Schüler:innen, etwa wenn diese gegenseitig ihre Bewertungen vergleichen, einen Eindruck von Intransparenz und fehlender Systematik in Bezug auf die Maßstäbe der Bewertung. Es besteht der berechtigte Anspruch, dass Bewertungskriterien einheitlich und vergleichbar sind. Es bleibt sowohl in den Bewertungsdokumenten als auch in der Benutzeroberfläche intransparent, ob es eine inhaltliche Bedeutung hat, wenn "Sprachstil" oder einfach nur "Stil", "Ausdruck" oder "Wortwahl", "Argumentationsfehler" oder "Unpräzise Argumentation" zurückgemeldet wird.

Die Auflistung von Fehlern, die keine sind, oder von widersprüchlichen Gesamtnoten sind ärgerliche Glitches, die – wenn sie so an Schüler:innen weitergereicht werden – nicht das Gefühl erzeugt, dass die Lehrkraft mit der gebotenen Sorgfalt und Ernsthaftigkeit arbeitet. So etwas wirkt demotivierend, zersetzend für die Beziehung zwischen Schüler:in und Lehrkraft, sowie delegitimierend für das Prinzip von Prüfung und Bewertung.

## 5. Die Umsetzung des Feedbacks führt nicht zur Verbesserung

In der Logik von Rückmeldungen auf Prüfungsleistungen oder Hausaufgaben ist angelegt, dass eine Umsetzung konkreter Verbesserungsvorschläge durch eine bessere Bewertung honoriert wird – oder jedenfalls nicht zu einer Verschlechterung der Bewertung führt (Monotonie). Die Auswertung von Testreihe B (Abbildung #fig:E:irrfahrt) zeigte hingegen, dass die vorgeschlagene Gesamtnote bei sukzessiver Umsetzung der Verbesserungsvorschläge des Korrekturtools nicht steigt, sondern in den ersten Iterationen sogar sinkt, während sie insgesamt um die Ursprungsbewertung oszilliert und dabei im Durchschnitt der 7–12 Iterationsstufen leicht *unterhalb* der Ursprungsbewertung verbleibt.

Diese Beobachtung gilt auch hinsichtlich des qualitativen Feedbacks. Betrachtet man das Feedback mehrerer Iterationen hintereinander, stellt man eine "Unentschiedenheit" oder "Flatterhaftigkeit" des Korrekturtools fest, welches die Schüler:innen abwechselnd in entgegengesetzte Richtungen weist; bei minutiöser Befolgung der Rückmeldungen werden die Änderungen im nächsten Durchlauf wieder zurück genommen.

Dieses Verhalten wirkt sich erheblich demotivierend auf Schüler:innen aus. Wenn auch über diverse Verbesserungsdurchläufe hinweg im Durchschnitt keine Verbesserung der Bewertung erkennbar ist, muss gefolgert werden, dass die inhaltlichen Rückmeldungen ohne didaktischen Wert sind. Das Prinzip von Feedback und Bewertung wird hiermit delegitimiert.

### 6. Erreichbarkeit einer Bestbewertung nur durch Einsatz von ChatGPT (Täuschung)

Schließlich ist es eine zentrale Beobachtung, dass in der Feedback-Logik des Korrekturtools nicht möglich zu sein scheint, eine perfekte – also nicht mehr zu beanstandende – Abgabe zu präsentieren. Wir konnten durch Umsetzungen der Rückmeldungen (und auch durch selbst geschriebene Lösungen) keine 99%-Bewertungen erreichen. Solche Bewertungen *konnten* allerdings erreicht werden, wenn die Lösung mithilfe von ChatGPT – dem LLM-basierten Chatbot, der selbst im Hintergrund des Korrekturtools verwendet wird – final überarbeitet wurde.

Die Botschaft an die Schüler:innen ist fatal: Lernen könnt ihr von dem Tool und seinen Rückmeldungen nichts; wenn ihr mit einer Bestbewertung abschneiden wollt, müsst ihr eure Abgabe von der KI selbst schreiben lassen.

## Gravität der Mängel und resultierende Forderungen an Fobizz und Anwender:innen

Tabelle #tab:D:gravität bietet eine Übersicht über die beschriebenen Mängel mit einer Einordnung, wie schwerwiegend sie jeweils im Hinblick auf die sachliche Eignung und ethische Vertretbarkeit des Einsatzes des Tools sind. Wir haben dazu ein dreistufiges Klassifikationssystem verwendet:

- **rot / "fatales Gebrauchshindernis":** Dies sind Mängel, die regelmäßig und in signifikanter Häufigkeit auftreten, gravierende Konsequenzen für die Schüler:innen (etwa in der Form inadäquater Bewertungen oder Rückmeldungen) bedeuten können, und die schwer zu erkennen oder auszumerzen sind.
- **gelb / "erhebliches Gebrauchshindernis":** Dies sind Mängel, die regelmäßig zu beobachten sind und das Prinzip von Prüfungsleistung und vergleichender, transparenter, fairer Bewertung unterminieren. Die geringfügig niedrigere Gravität resultiert daher, dass die Mängel entweder bei manueller Durchsicht mit normalem Aufwand erkennbar sind oder keine direkte Auswirkung auf die Benotung haben.
- **blau / "Erschwernis des Gebrauchs":** Dies sind Mängel, die leicht erkennbare "Glitches" darstellen oder sich mehr oder weniger darauf beschränken, dass die Fähigkeiten des Tools irreführend kommuniziert wurden.

Des Weiteren haben wir in der Tabellenspalte "Prinzipielle Mangel oder technisch lösbar?" eine grobe und erfahrungsbasierte Einordnung in Bezug auf die mutmaßliche Ursache und Behebbarkeit der Mängel vorgenommen. Zu beachten ist, dass die meisten der hier festgestellten Mängel typische und wissenschaftlich dokumentierte Eigenschaften von großen Sprachmodellen (LLMs) wie GPT-4 widerspiegeln. In welcher Weise zukünftige technologische Innovationen diese Mängel nicht mehr aufweisen, ist naturgemäß eine spekulative Frage. Bei der Unterscheidung, ob es sich um prinzipielle oder technisch behebbare Mängel handelt, haben wir uns daher am aktuellen Stand der Technik orientiert und – da wir keinen Zugang zum Quellcode der Beteiligten Algorithmen haben – geschulte Schätzungen unternommen. Dabei haben wir ein dreistufiges Klassifikationssystem verwendet:

- **rot / "prinzipiell":** Dies sind Mängel, die auf fundamentale Eigenschaften von LLMs zurück gehen und bei denen wir absehbar keine Hoffnung haben, dass sich diese technisch beheben lassen.
- **gelb / "durch bessere Programmierung vermutlich in der Häufigkeit reduzierbar"** nennen wir jene Mängel, die ebenfalls auf prinzipielle Eigenschaften von LLMs zurück gehen, bei denen wir jedoch erfahrungsbasiert *vermuten*, dass eine sorgfältigere Programmierung / Prompting oder die Verwendung fortgeschrittener Techniken wie prompt chaining[44] den Mangel zumindest in der Frequenz deutlich reduzieren könnten.

[44] https://www.ibm.com/de-de/topics/prompt-chaining ; https://www.promptingguide.ai/techniques/prompt_chaining

- **blau / “vermutlich durch bessere Programmierung behebbar”** nennen wir jene Mängel, bei denen wir vermuten, dass sie sich durch verfügbare Programmiertechniken mehr oder weniger gänzlich in den Griff bekommen ließen.

Die fünfte Tabellenspalte enthält schließlich eine Forderung an Fobizz und/oder die Anwender:innen der Software. Diese Forderungen verstehen sich als Schlussfolgerungen aus dem gezeigten Mangel, seiner diskutierten Gravität und unserer Einschätzung über seine Ursache bzw. Behebbarkeit. Diese Schlussfolgerungen beruhen auf folgenden Grundsätzen:

- Stellt ein Mangel ein erhebliches oder fatales Gebrauchshindernis dar *und* ist er prinzipieller Natur (d.h., beruht auf einer fundamentalen Eigenschaft von LLMs), dann ist nach unserem Dafürhalten die Schlussfolgerung, dass das Tool nicht angeboten, bereitgestellt und verwendet werden sollte. Diese Schlussfolgerung erreichen wir in den Fällen 1.a, 1.b, 2.a und 5.a/b.
- Lässt sich ein erheblicher Mangel zwar prinzipiell nicht beheben, aber in seiner Häufigkeit deutlich reduzieren, ist für uns die Schlussfolgerung, das Tool so lange nicht anzubieten oder zu verwenden, wie dies nicht geschehen ist (siehe 2.b).
- In den Fällen 3.a-c, 4.a-c, 6 bestehen unsere Schlussfolgerungen in konkreten Verbesserungsvorschlägen, die entweder eine Beschränkung des Einsatzspektrums des Tools oder einen transparenten und proaktiven Umgang mit den Limitation des Tools fordern.

| **Mangel** | **Beschreibung** | **Gravität**<br><br>Zusätzliche Beobachtungen | **Prinzipieller Mangel oder technisch lösbar?** | **Resultierende Forderungen an Fobizz/Anwender:innen** |
|---|---|---|---|---|
| 1.a | Zufälligkeit der vorgeschlagenen Gesamtnote | **Fatales Gebrauchshindernis**<br><br>erschwerte Erkennbarkeit des Mangels durch Nutzende | **prinzipiell** | **Tool nicht anbieten/verwenden** |
| 1.b | Zufälligkeit der inhaltlichen Bewertung und qualitativen Rückmeldung | **Fatales Gebrauchshindernis**<br><br>erschwerte Erkennbarkeit des Mangels durch Nutzende | **prinzipiell** | **Tool nicht anbieten/verwenden** |
| 2.a | Unzuverlässige Erkennung von Falschbehauptungen | **Erhebliches Gebrauchshindernis**<br><br>Unterminierung des Wertversprechens, Delegitimierung von Prüfung und Benotung | **prinzipiell** | **Tool nicht anbieten/verwenden** |
| 2.b | Unzuverlässige Erkennung von Nonsense und Arbeitsverweigerung | **Erhebliches Gebrauchshindernis**<br><br>Unterminierung des Wertversprechens, Delegitimierung von Prüfung und Benotung | durch bessere Programmierung vermutlich in der Häufigkeit reduzierbar | Tool nicht anbieten/verwenden, solange der Mangel in seiner Häufigkeit nicht auf weniger “corner cases” beschränkt ist. |
| 3/a/b | Unzuverlässige Umsetzung einzelner | **Erschwernis des Gebrauchs** | **prinzipiell** | Beschränkung des Tools auf beherrschbare Kriterien; |
| | Bewertungskriterien – z.B. a) Textlänge, b) Erkennung KI-generierter Abgaben | Irreführung in Bezug auf die Fähigkeiten des Tools | | Transparenz und Aufklärung auf Ebene von Benutzerführung und Bewertungsdokumenten. |
| 4.a | a. Uneinheitliche und zufällige Bezeichnung von Fehlerkategorien | **Erhebliches Gebrauchshindernis**<br><br>mangelnde Transparenz und Einheitlichkeit der Bewertungsmaßstäbe; Delegitimierung von Prüfung und Benotung | vermutlich durch bessere Programmierung behebbar | Verbesserung des Tools. Transparenz und Aufklärung auf Ebene von Benutzerführung und Bewertungsdokumenten. |
| 4.b/c | Inkonsistentes Feedback durch a) Angabe von Fehlern, die keine sind; b) Widersprüchliche Angabe der Gesamtnote | **Erschwernis des Gebrauchs**<br><br>Irreführung in Bezug auf die Fähigkeiten des Tools, Motivationshemmnis und Irreführung der Schüler:innen | durch bessere Programmierung vermutlich in der Häufigkeit reduzierbar | Glitches erkennen und filtern. |
| 5.a/b | Die Umsetzung des Feedbacks führt nicht zur Verbesserung | **Fatales Gebrauchshindernis**<br><br>Mangel didaktischer Fundierung; Delegitimierung von Prüfung und Benotung | **prinzipiell** | **Tool nicht anbieten/verwenden** |
| 6 | Erreichbarkeit einer Bestbewertung nur durch Einsatz von ChatGPT (Täuschung) | **Fatales Gebrauchshindernis**<br><br>Mangel didaktischer Fundierung; Delegitimierung von Prüfung und Benotung | durch bessere Programmierung vermutlich in der Häufigkeit reduzierbar | Tool nicht anbieten/verwenden, solange der Mangel in seiner Häufigkeit nicht auf weniger "corner cases" beschränkt ist. |

**Tab. #tab:D:gravität:** Vergleich und Einordnung der beobachteten funktionalen Defizite mit Blick darauf, wie schwer sie wiegen (Spalte 3), ob technische Lösungen dafür erwartbar sind (Spalte 4) und welche Schlussfolgerungen für Anbieter und Abnehmer des Tools wir daraus ableiten (Spalte 5).

## Weiterführende Untersuchungen

Unsere Untersuchung stellt mit einem Stichprobenumfang von 1 exemplarischen Aufgabenstellung, 10 simulierten Schülerabgaben und 50 getesteten Korrekturläufen des Korrekturtools in Testreihe A eine qualitative Untersuchung mit quantitativer Evidenz dar. Die 10 Abgaben und 5 Korrekturläufe pro Abgabe halten wir für qualitativ hinreichend aussagekräftig. Von Interesse für nachfolgende Untersuchungen wäre besonders eine Variation der Aufgabenstellung hinsichtlich Komplexität, Fachzugehörigkeit und Anforderungsprofil. Unsere Aufgabenstellung ist vergleichsweise simpel und kann mit Allgemeinwissen gut bearbeitet werden. Insbesondere mit Blick auf Aufgabenstellungen, die längere und komplexere Abgaben zur Folge hätten (z.B. mehrseitige Aufsätze), Textinterpretationen, die Arbeit mit historischen Quellen, oder fiktionales Schreiben umfassen würden, sind unsere Tests weniger aussagekräftig. Auf der Grundlage stichprobenartiger Tests erwarten wir, dass das Tool bei Aufgabe mit höherer Komplexität nicht besser sondern eher noch schlechter abschneidet.

In Bezug auf Testreihe B, in der wir zunächst nur 2 Texte überprüft haben, um qualitative Evidenz zu gewinnen, werden wir in einer erweiterten Version dieser Studie den Stichprobenumfang deutlich erhöhen. Diese Untersuchungsmethode ist sodann auch auf das im November 2024 neu veröffentliche “KI-Feedbacktool” von Fobizz anzuwenden, welches im Gegensatz zur KI-Korrekturhilfe für die Benutzung durch Schüler:innen selbst angeboten wird.

Ebenfalls wäre eine Variation der Bewertungskriterien und ein ausführlicheres Experimentieren mit dem Eingabefeld “Sample Solution or Scope” von Interesse, um zu ermitteln, wie sich die Performance des Tools dadurch verändert. In Bezug hieraus sind unsere Erwartungen allerdings aus zwei Gründen zurückhaltend: Erstens gehen die Meisten der gezeigten Mängel auf prinzipielle Eigenschaften von LLMs zurück und lassen sich dadurch nicht durch geschickteres Prompting über die Eingabebox “Sample Solution or Scope” verhindern. Zweitens wird das Tool in der Bewerbung und im Nutzerinterface nicht so kommuniziert, dass “geschicktes Prompting” für seinen adäquaten Gebrauch erforderlich sei – und es würde dem Wertversprechen entgegenlaufen, wenn nur GPT-4-versierte Nutzer:innen es verantwortungsvoll verwenden könnten. Wir haben uns mit unserem Test also bewusst an der mutmaßlichen Verwendungsweise einer “allgemeinen” Nutzer:in orientiert.

Wir haben zu Beginn von Abschnitt II darauf hingewiesen, dass wir die Leistung des Korrekturtools hier nicht vor dem Hintergrund eines normativen didaktischen Diskurses über gute und angemessene Rückmeldung und Benotung beurteilt haben. Wir haben die Leistung und Tauglichkeit vielmehr im Hinblick auf so gravierende funktionale Defizite hin untersucht, dass diese Mankos objektive Gebrauchshindernisse darstellen. Als Implikation dieser in Bezug auf die didaktische Theorie agnostischen Herangehensweise haben wir nicht untersucht, was das Tool gut macht, sondern nur fatale Gebrauchshindernisse benannt. Dieser negative Fokus mag einseitig erscheinen. Eine Abwägung von didaktischer Vor- und Nachteilen kommt aus unserer Sicht allerdings nur für ein Tool in Betracht, das die hier gelisteten Mankos nicht aufweist, durch die das Tool die Mindestschwelle für eine wirklich didaktische Beurteilung gar nicht erreicht.[45]

# V. Fazit

Das Lesen, Kommentieren und Bewerten von Aufsätzen kostet Zeit, Geduld und Besonnenheit. Das Versprechen einer Zeitersparnis in diesem Bereich dürfte deshalb einen neuralgischen Punkt bei vielen über Monate an ihrer Belastungsgrenze arbeitenden Lehrkräften treffen. Aber lässt sich diese Tätigkeit wirklich automatisieren? Immerhin umfasst Bewerten und Beurteilen mehr, als Rechtschreibfehler zu finden – was jedes Textverarbeitungsprogramm durch einen mechanischen Abgleich mit einem Wörterbuch erledigen kann. Bewerten und Beurteilen erfordert in einem umfassenderen Sinne menschliches Urteilsvermögen, didaktische Expertise und zwischenmenschliches Feingefühl. Aus all diesen Gründen ist die Automatisierung von Beurteilung und Bewertung ein mit besonderer Vorsicht zu genießendes Wertversprechen auf dem Markt für KI-Angebote für Lehrkräfte. All dies motivierte uns dazu, aus der Angebotspalette von Fobizz speziell die “KI-Korrekturhilfe” unter die Lupe zu nehmen.

## 1. Die “KI-Korrekturhilfe” von Fobizz sollte im Schulalltag nicht eingesetzt werden

Unsere empirische Untersuchung hat sich auf **funktionale Mindestbedingungen für die Gebrauchstauglichkeit** der Fobizz “KI-Korrekturhilfe” fokussiert und mehrere gravierende Beeinträchtigungen gefunden (siehe Tab. #tab:D:gravität und Abschnitt III). Zu den gravierendsten Mängeln gehören:

[45] In einer relevanten ähnlichen Studie haben Kolleg:innen die Qualität der numerischen Bewertungen von Deutsch-Aufsätzen durch LLM-Modelle untersucht, indem sie die automatisierten Bewertungen mit den durch Leher:innen vergebenen Bewertungen verglichen haben. Siehe https://arxiv.org/abs/2411.16337.

**(1)** eine starke Zufallsabhängigkeit der numerischen Bewertung und der qualitativen Rückmeldung, was erst bei mehrfacher Wiederholung der automatisierten Bewertung für dieselbe Lösung erkennbar ist.

**(2)** Falschbehauptungen und Nonsense-Lösungen werden nicht zuverlässig erkannt, weshalb die Verwendung des Tools leicht einer Abwertung und Diskreditierung des Prinzips von Prüfung und Bewertung Vorschub leistet.

**(3+4)** Das Tool weist eine unzureichende Transparenz im Umgang mit den Bewertungskriterien und seinen eigenen Einschränkungen auf. So lassen sich Bewertungskriterien einstellen, die vom Tool nicht zuverlässig verarbeitet werden können, ohne dass dies transparent gemacht wird. Dadurch entsteht ein irreführender Eindruck sowohl bei den Nutzer:innen als auch bei den Empfänger:innen der Rückmeldungen hinsichtlich des tatsächlich angewendeten Bewertungssystems.

**(5)** Die Umsetzung der Verbesserungsvorschläge des Tools führt nicht zu einer konsistenten Verbesserung der Bewertung; im Durchschnitt sinkt die Gesamtnote bei iterativer Verbesserung einer Abgabe, was die inhaltliche Qualität des Feedbacks und der Bewertungsprinzipien delegitimiert.

**(6)** Eine Bestbewertung konnten wir nur mit Abgaben erreichen, die mithilfe des Chatbotsystems ChatGPT von OpenAI verfasst oder überarbeitet wurden; dieses Verhalten regt Schüler:innen zu Täuschungsversuchen an.

Diese Mängel sind vor dem Hintergrund der wissenschaftlich dokumentierten technischen Eigenschaften großer Sprachmodelle (LLMs) *nicht* erstaunlich. So lassen sich alle diese Mängel mit den typischen Verhaltenseigenschaften des im Hintergrund verwendeten Modells GPT-4 gut erklären. Deshalb ist eine **technische Lösung für die Mängel nicht in Sicht** und eine grundlegende Verbesserung des Tools in einer nächsten Version nicht erwartbar; lediglich eine Reduktion der Häufigkeit einiger der beobachteten Probleme scheint uns hier und da durch verbesserte Programmierung (“prompting”) in Aussicht zu stehen.

Ausgehend von dieser grundlegenden technischen Einschätzung von LLMs und ihren Limitationen ist der Kern unserer Kritik an Fobizz die **unverantwortliche Gestaltung (Benutzerführung) und Vermarktung des Korrekturtools.** Ein transparenter und realistischer Umgang mit den eigenen Begrenzungen wäre ein Muss für eine seriöse Kommunikation, die sich der Verantwortung auch für die Schüler:innen, deren Bildungswege von einzelnen Notenentscheidungen abhängen können, im Mindesten gerecht zu werden. Besonders von einem Unternehmen, das neben Softwaretools als Anbieter von Lehrkräfteschulungen erfolgreich ist und Vertrauen genießt, wäre zu verlangen, dass es über die Grenzen der Gebrauchstauglichkeit eines solchen Produkts proaktiv aufklärt.

Lehrkräften, die dieses Tool möglicherweise verwenden möchten, müssen wissen, dass ein gewinnbringender und verantwortungsbewusster Gebrauch **ein großes Wissen über LLMs voraussetzt.** Diese Voraussetzung ist unrealistisch für ein Produkt, dessen Wertversprechen in der Zeit- und Arbeitsersparnis für die Lehrkräfte besteht und somit potenziell in Situationen von Stress und Überforderung zum Einsatz kommt. Das wissenschaftlich bekannte Phänomen des *automation bias* (Automatisierungsvoreingenommenheit) wird im Gegenteil sogar bewirken, dass zahlreiche Anwender:innen den Ausgaben des Tools (numerischen Bewertungen und inhaltlichen Rückmeldungen) *mehr* Vertrauen schenkt, als einer menschlichen Bewertung.

Als unabdingbare **Sofortmaßnahme** ist allen Lehrkräften, die dieses Tool einsetzten, zu raten, jede Bewertung, die sie damit erstellen, mehrfach (3–5 mal) generieren zu lassen und die Resultate zu vergleichen, um ein Gefühl für die Zufallsabhängigkeit der Ausgabe zu erhalten. Die Vergabe der numerischen Noten sollte ausschließlich durch die Lehrkraft selbst auf Grundlage didaktischer Kriterien und eines eigenen Urteils erfolgen. Sämtliche inhaltlichen Rückmeldungen, die das Tool automatisiert erstellt, müssen einem “fact checking” unterzogen werden; außerdem müssen die Lösungen auf inhaltliche Fehler und Falschbehauptungen hin durchgelesen werden, die das Tool häufig übersieht. Lehrkräfte sollten nicht erwarten, dass das Tool didaktisch fundierte und konsistente Rückmeldungen gibt, Schüler:innen werden durch oszillierendes und unentschiedenes Feedback “im Kreis geschickt”.

## 2. KI in der Schule ist keine Lösung für den Lehrkräftemangel und Produkt eines gefährlichen gesellschaftlichen Trends

Unsere Untersuchung nimmt nur ein exemplarisches KI-Tool in den Blick. Mit der Automatisierung von Bewertung und Rückmeldung möchte dieses exemplarische Tool ein zentrales Wertversprechen LLM-basierter KI-Systeme für den Bildungskontext einlösen. Gerade weil viele der beobachteten Mängel allerdings typisch für LLM-Systeme im Allgemeinen sind, ist es an dieser Stelle sinnvoll, den Blick wieder auf den gesellschaftspolitischen Kontext zu weiten und einige Verallgemeinerungen aus den exemplarischen Beobachtungen zu extrapolieren.

Die gesellschaftliche Situation, in der KI-Tools für den Einsatz in der Schule angeschafft werden, ist – wie in Abschnitt I geschildert – durch Lehrkräftemangel, eine chronische Überlastung des Bildungssystems und der vorhandenen Lehrkräfte, sowie unzureichende Finanzierung des Bildungssystems gekennzeichnet. In dieser Situation ist der Ruf nach künstlicher Intelligenz als schnelle Lösung der typische Fall eines “Techno Fix” im Sinne des “Techno Solutionism”.[46] Denn hiermit erfolgt die Verschiebung eines sozialen und politischen Problems in die Sphäre der Technologie – sozialpolitisches Versagen soll mittels Technik gelöst werden. Eine Industrie und spezialisierte Geschäftsmodelle stehen bereit, um solchen Gemengelagen Kapital zu schlagen, indem sie den oftmals weniger fachkompetenten politischen Entscheidungsträger:innen vordergründig innovative, aber ungetestete Produkte in einem experimentellen Entwicklungsstadium verkaufen. Der Fall der Fobizz „KI-Korrekturhilfe“ lässt sich mit einem Automobilhersteller vergleichen, der ein Auto ohne Bremsen und Rückspiegel auf den Markt bringt und sich erst nachträglich um Sicherheitstests und eine mögliche TÜV-Zulassung kümmert.

Das Bundesland Rheinland-Pfalz hat für seine Fobizz-Landeslizenz mit einer Laufzeit von Dezember 2023 bis Juli 2025 einen Betrag von 1.77 Mio € netto gezahlt.[47] Dasselbe Bundesland war im Sommer 2024 mit der Nachricht in den Schlagzeilen, seine ca. 2.000 Lehramtsreferendar:innen über die Sommerferien vor ihrer Übernahme in den Schuldienst nicht zu bezahlen – sie also für sechs Wochen in die Arbeitslosigkeit zu entlassen.[48] Die Unattraktivität des Lehrerberufs ist ein Hauptgrund für den schlechten Stand des Schulsystems. Ein “schonender Umgang mit Steuergeldern”, den ein Sprecher des rheinland-pfälzischen Bildungsministeriums in der Berichterstattung als Grund für die Nichtbezahlung der Referendar:innen anführt,[49] ist bei der Beschaffung einer Fobizz-Landeslizenzen nicht so leicht erkennbar.

Die technische Lösung sozialpolitischer Probleme mittels schlecht evaluierter KI-Tools ist allerdings nicht nur möglicherweise eine Verschwendung von Steuergeldern. Sie ist darüber hinaus gefährlich, denn existenzielle Entscheidungen über Menschen, die typischerweise wenig Ressourcen haben, sich zu wehren oder die Entscheidungen anzufechten, werden hier leichtfertig einer Kombination von blinder Automatisierung und Profitnahme der beteiligten Unternehmen unterworfen. So ist aus zahlreichen Gesellschaftsbereichen – vom Finanz- und Versicherungswesen über Personalführung, Schul- und Studienplatzvergabe, bis zu Vergabeentscheidungen bei Sozialhilfe und Transferleistungen, in der Altenpflege, dem Jugendschutz und der Polizeiarbeit bekannt und gut dokumentiert, dass der Einsatz von KI-Tools eine neue, verschärfte Formen der Ungerechtigkeit durch

[46] Morozov, Evgeny. 2013. To Save Everything, Click Here: The Folly of Technological Solutionism. New York: PublicAffairs.
[47] Persönliche Korrespondenz mit dem Ministerium für Bildung, Rheinland-Pfalz.
[48] https://www.zdf.de/nachrichten/politik/deutschland/referendariat-lehramt-sommerferien-100.html
[49] ebd.

Stereotypisierung, Biases und Fehleranfälligkeit der Tools verursacht.[50] Durch die Verlagerung funktionaler Kompetenzen von öffentlich-rechtlich beschäftigtem Personal sozialer Einrichtungen an privatwirtschaftliche Unternehmen (und im Fall von Fobizz stehen hier internationale Großunternehmen wie OpenAI und Microsoft im Hintergrund) machen sich Bildungs- und soziale Fürsorgesysteme abhängig von Kapitalinteressen.[51]

KI-Tools können zweifellos einen positiven Nutzen für Lehrkräfte bieten. Derzeit befinden sich diese Technologien jedoch noch in einem experimentellen Stadium, und ihr systematischer Einsatz erfordert spezifische technische Kompetenzen seitens der Lehrkräfte. Schulträger, Bildungsministerien und Fachdidaktiker:innen stehen vor der anspruchsvollen Aufgabe, zuverlässige Evaluations- und Akkreditierungsverfahren für diese Tools zu entwickeln und umzusetzen. Gesellschaftspolitisch besteht die Herausforderung darin, die Hoffnung auf KI im Schulkontext nicht dazu zu nutzen, die grundlegenden Probleme des Bildungssystems zu kaschieren. Statt auf verstärkte Automatisierung in diesem sensiblen sozialen Bereich zu setzen, sind politisches Handeln und umfassende Investitionen in das Bildungssystem erforderlich.

**Materialanhang**
Der Materialanhang ist erhältlich unter: https://rainermuehlhoff.de/Fobizz-Paper-2024-Materialanhang/

**Acknowledgements**
Wir danken unseren studentischen Mitarbeitenden, die an dieser Studie mitgewirkt haben. Melissa Schnabel hat die Durchführung und Protokollierung der Testreihen unterstützt. Elena Herold und Hedye Tayebi-Jazayeri haben zahlreiche Hintergrundrecherchen vorgenommen.

# VI. Update 21.01.2025: Reaktion des Unternehmens Fobizz

Nach der Veröffentlichung unserer Studie am 09.12.2024 auf der Preprint-Plattform arXiv hat das Unternehmen Fobizz auf verschiedene Weise reagiert.

**Als experimentell getagged:** Die "KI-Korrekturhilfe" wurde im internen Auswahlmenü auf fobizz.com mit einem roten Tag "experimental" versehen.

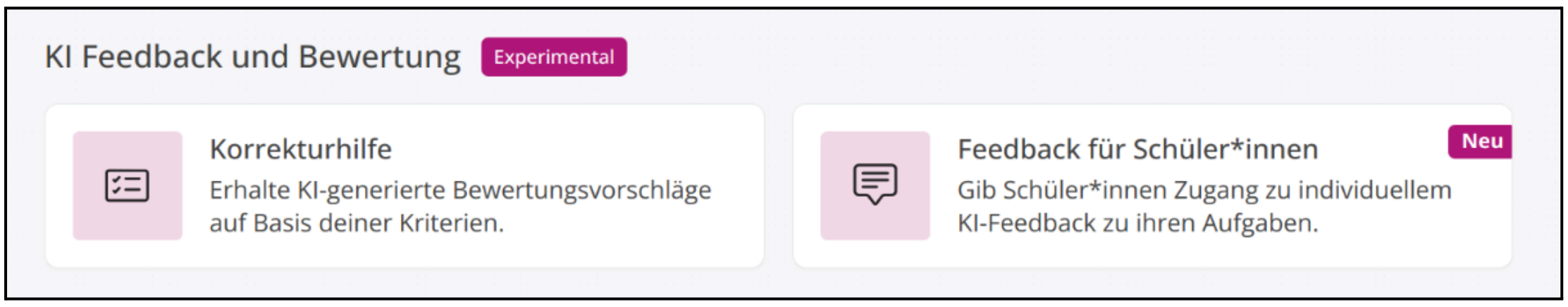

Screenshot 2024-12-18: https://tools.fobizz.com/ (nach Login)

[50] Eubanks, Virginia. 2017. *Automating Inequality: How High-Tech Tools Profile, Police, and Punish the Poor*. New York, NY: St. Martin's Press.
O'Neil, Cathy. 2017. *Angriff der Algorithmen: Wie sie Wahlen manipulieren, Berufschancen zerstören und unsere Gesundheit gefährden*. Translated by Karsten Petersen. 2nd ed. Carl Hanser Verlag GmbH & Co. KG.
Mühlhoff, Rainer. 2020. "Automatisierte Ungleichheit: Ethik der Künstlichen Intelligenz in der biopolitischen Wende des Digitalen Kapitalismus." *Deutsche Zeitschrift für Philosophie* 68 (6): 867–90. https://doi.org/10.1515/dzph-2020-0059.
[51] Zu diesen Interessen gehört zum Beispiel die Wiederverwendung der bei der Nutzung der Tools anfallenden Daten: Im Fall von digitalen Produkten in der Bildung ist Learning Analytics (eine Technik, die auch die automatisierte Profilbildung über Lernende umfasst) ein profitabler und wachsender Industriezweig. Eine Sekundärverwendung der Daten, die im Rahmen von KI-Lösungen für Lehrer:innen oder Schüler:innen anfallen, für Profilanalysen kann auch anonymisiert und somit datenschutzkonform erfolgen. Vgl.:
Mühlhoff, Rainer. 2024. "Das Risiko Der Sekundärnutzung Trainierter Modelle Als Zentrales Problem von Datenschutz Und KI-Regulierung Im Medizinbereich." In *Der Einsatz von KI & Robotik in Der Medizin*, edited by Hannah Ruschemeier and Björn Steinrötter, 27–52. Nomos. https://doi.org/10.5771/9783748939726-27.
Mühlhoff, Rainer. 2023. *Die Macht der Daten: Warum künstliche Intelligenz eine Frage der Ethik ist*. V&R unipress, Universitätsverlag Osnabrück. https://doi.org/10.14220/9783737015523.

**Disclaimer, der einige unserer Mängelpunkte benennt:** Das Unternehmen hat die Hilfe-Seite zur KI-Korrekturhilfe um zwei Abschnitte “Was noch nicht zuverlässig funktioniert” und “Was noch gar nicht funktioniert” erweitert. Die hier aufgeführten fünf Punkte geben einige der in dieser Studie vorgebrachten Mängelpunkte wieder. Wir weisen darauf hin, dass Mangel 5 (Umsetzung des Feedbacks führt nicht zur Verbesserung) nicht dadurch behoben werden muss, dass das Korrekturtool “Kenntnis von vorherigen Versionen” behält, wie es die Hilfeseite suggeriert. Wir würden erwarten, dass die Umsetzung von Verbesserungsvorschlägen bei einer Neubewertung ohne Kenntnis der Vorversion zumindest nicht zu einer Verschlechterung führt.

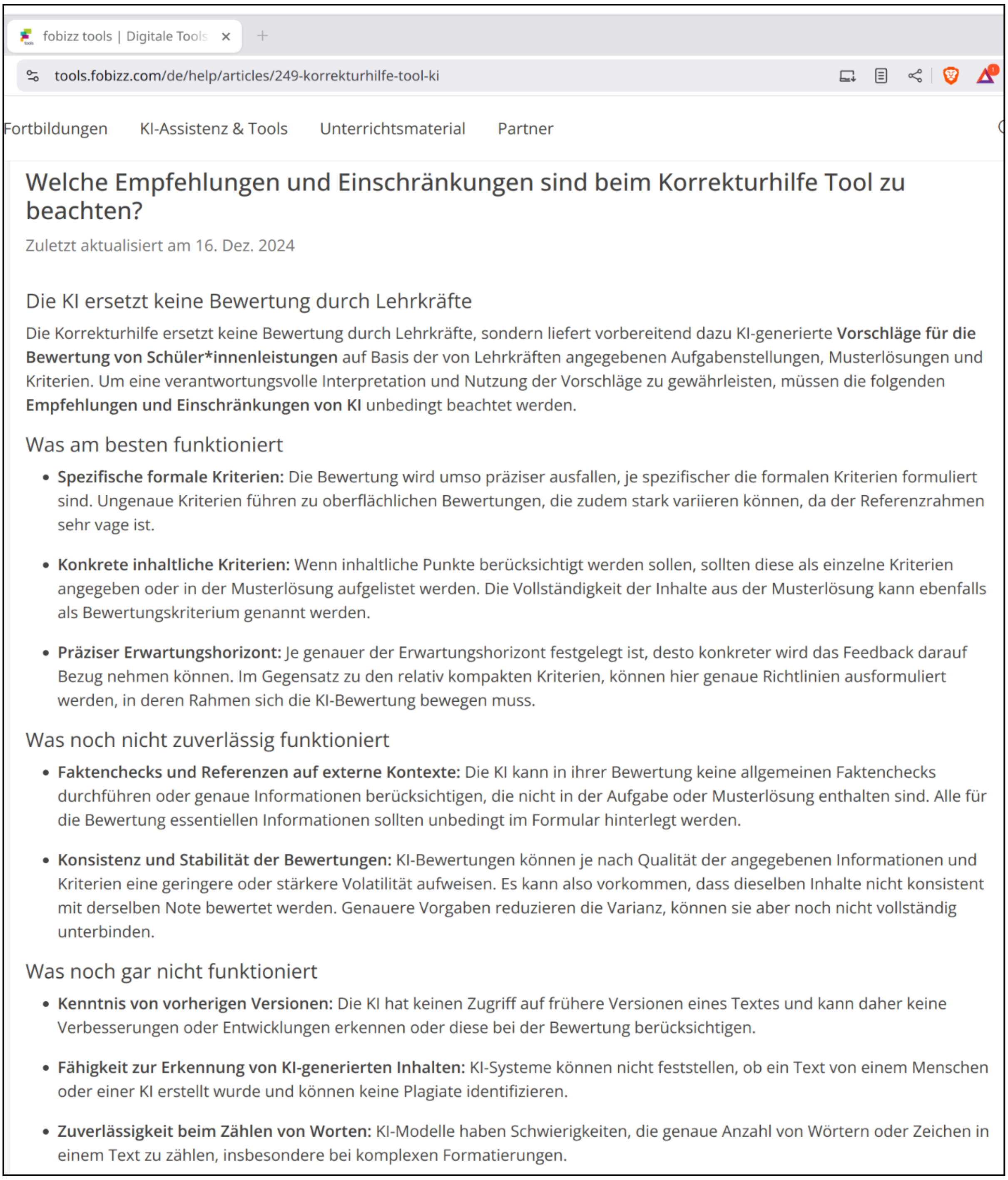
fobizz tools | Digitale Tools

tools.fobizz.com/de/help/articles/249-korrekturhilfe-tool-ki

Fortbildungen KI-Assistenz & Tools Unterrichtsmaterial Partner

## Welche Empfehlungen und Einschränkungen sind beim Korrekturhilfe Tool zu beachten?

Zuletzt aktualisiert am 16. Dez. 2024

### Die KI ersetzt keine Bewertung durch Lehrkräfte

Die Korrekturhilfe ersetzt keine Bewertung durch Lehrkräfte, sondern liefert vorbereitend dazu KI-generierte **Vorschläge für die Bewertung von Schüler*innenleistungen** auf Basis der von Lehrkräften angegebenen Aufgabenstellungen, Musterlösungen und Kriterien. Um eine verantwortungsvolle Interpretation und Nutzung der Vorschläge zu gewährleisten, müssen die folgenden **Empfehlungen und Einschränkungen von KI** unbedingt beachtet werden.

### Was am besten funktioniert

- **Spezifische formale Kriterien:** Die Bewertung wird umso präziser ausfallen, je spezifischer die formalen Kriterien formuliert sind. Ungenaue Kriterien führen zu oberflächlichen Bewertungen, die zudem stark variieren können, da der Referenzrahmen sehr vage ist.
- **Konkrete inhaltliche Kriterien:** Wenn inhaltliche Punkte berücksichtigt werden sollen, sollten diese als einzelne Kriterien angegeben oder in der Musterlösung aufgelistet werden. Die Vollständigkeit der Inhalte aus der Musterlösung kann ebenfalls als Bewertungskriterium genannt werden.
- **Präziser Erwartungshorizont:** Je genauer der Erwartungshorizont festgelegt ist, desto konkreter wird das Feedback darauf Bezug nehmen können. Im Gegensatz zu den relativ kompakten Kriterien, können hier genaue Richtlinien ausformuliert werden, in deren Rahmen sich die KI-Bewertung bewegen muss.

### Was noch nicht zuverlässig funktioniert

- **Faktenchecks und Referenzen auf externe Kontexte:** Die KI kann in ihrer Bewertung keine allgemeinen Faktenchecks durchführen oder genaue Informationen berücksichtigen, die nicht in der Aufgabe oder Musterlösung enthalten sind. Alle für die Bewertung essentiellen Informationen sollten unbedingt im Formular hinterlegt werden.
- **Konsistenz und Stabilität der Bewertungen:** KI-Bewertungen können je nach Qualität der angegebenen Informationen und Kriterien eine geringere oder stärkere Volatilität aufweisen. Es kann also vorkommen, dass dieselben Inhalte nicht konsistent mit derselben Note bewertet werden. Genauere Vorgaben reduzieren die Varianz, können sie aber noch nicht vollständig unterbinden.

### Was noch gar nicht funktioniert

- **Kenntnis von vorherigen Versionen:** Die KI hat keinen Zugriff auf frühere Versionen eines Textes und kann daher keine Verbesserungen oder Entwicklungen erkennen oder diese bei der Bewertung berücksichtigen.
- **Fähigkeit zur Erkennung von KI-generierten Inhalten:** KI-Systeme können nicht feststellen, ob ein Text von einem Menschen oder einer KI erstellt wurde und können keine Plagiate identifizieren.
- **Zuverlässigkeit beim Zählen von Worten:** KI-Modelle haben Schwierigkeiten, die genaue Anzahl von Wörtern oder Zeichen in einem Text zu zählen, insbesondere bei komplexen Formatierungen.

Screenshot 2024-12-18: https://tools.fobizz.com/de/help/articles/249-korrekturhilfe-tool-ki
Die letzten beiden Abschnitte sind hinzugefügt worden und übernehmen unsere Mängelpunkte.

**High level meeting mit Fobizz:** Am 19.12.2024 kam es zu einem Treffen zwischen Vorstandsmitgliedern und Produktmanagement von Fobizz und unserem Team. Fobizz war sehr an einem konstruktiven Dialog interessiert. In der ausführlichen Unterredung ging es um unternehmerische Verantwortung und den Übergang von einem Start-Up, das “einfach mal Produkte anbietet und dann schaut, was für Feedback kommt”, und einem Unternehmen mit Bundesland-Lizenverträgen, dessen Tools steuergelderfinanziert allen Lehrkräften zugänglich gemacht werden. Wichtigste direkte Erkenntnisse für uns:

- Automatisierte Korrekturen seien eines der “meistgewünschten KI-Tools” im Repertoire der Firma.
- Das Tool werde seit etwa einem Jahr angeboten.
- Das Tool sei bis jetzt nicht systematisch getestet worden.

  Im Gesprächsverlauf haben wir die Veröffentlichung ungetester Korrekturtools als unverantwortlich herausgestellt und die Notwendigkeit systematische Testung solcher Tools betont, da für Schüler:inne mutmaßlich einiges von solchen Rückmeldungen und Bewertungen abhängt.
- Der hinter dem Tool liegende ChatGPT-Prompt sei bislang weniger als eine halbe DIN-A4-Seite lang gewesen. Man habe aber nun eine neue Version des Tools erstellt, dessen Promt ca. 9 Seiten umfasse. Der erweiterte Prompt wurde uns am 19.12.2025 zur Einsicht zur Verfügung gestellt.
- Der Vorstand schlägt vor, dass wir unsere Tests mit der geupdateten Version des Korrekturtools wiederholen könnten. (Wir stellen in Aussicht, das Tool weiterhin zu testen, committen uns aber nicht bezüglich der Timeline oder konkreten Teststrategie.)

**Der neue Prompt enthält unseren Testcase als Beispiel:** Das überarbeitete und nach Auskunft des Unternehmens seit 15.12.2024 im Einsatz befindliche Korrekturtool basiert auf einem deutlich erweiterten Prompt, der uns zur Einsicht überlassen wurde. Wie üblich beim Erstellen von ChatGPT-Prompts enthält dieser auch Beispiele, bestehend aus Aufgabenstellungen, Bewertungskriterien und Abgabe der Schüler:in. Solche Beispiele bilden einen Teil der Instruktionen an das LLM-basierte KI-System, nach dem dieses seine “Rolle” ausfüllen soll – hier die Rolle einer “KI-Assistenz, die Lehrkräfte dabei unterstützt, die Ergebnisse ihrer Schüler:innen […] zu bewerten”.

Der erweiterte Prompt enthält konkret *zwei* Beispiele, von denen das zweite auffallende Ähnlichkeiten mit unserem Testszenario aufweist. Die Aufgabenstellung und die Bewertungskriterien wurden mit leichten Abwandlungen übernommen, wie die folgende Gegenüberstellung zeigt:

| **Unser Test-Szenario (Testreihe A)** | **Beispiel aus dem Fobizz-Prompt vom 15.12.2024** |
|---|---|
| **## Assignment**<br>Schreibe eine begründete Stellungnahme für oder gegen die Absenkung des Wahlalters auf 14 Jahre. Gehe dabei auf mindestens ein Argument für jede Seite ein und beziehe anschließend Position.<br>(Schreibe ca. 150 - 250 Wörter.)<br><br>**## Solution**<br>Erwartungshorizont:<br>- 10te Klasse<br>- Einleitungssatz<br>- Mind. ein Argument dafür. | **<Beispiel 2>**<br>**## Assignment**<br>Verfasse eine begründete Stellungnahme, in der du entweder die Einführung von Schuluniformen an deiner Schule befürwortest oder ablehnst. Berücksichtige dabei mindestens ein Argument für jede Position und teile am Ende deine eigene Meinung mit.<br><br>**## Solution**<br>Erwartungshorizont für die Stellungnahme zur Einführung von Schuluniformen<br>Klassenstufe: 10. Klasse<br>Thema: Einführung von Schuluniformen an der Schule<br>1. Einleitungssatz: |
| - Mind. ein Argument dagegen.<br>- Klare Stellungnahme für eine der beiden Seiten.<br>- Stellungnahme passt zu den Argumenten.<br>- Die vertretene Position lässt sich von den Argumenten ableiten.<br>- Argumentation ist logisch schlüssig.<br>- Beinhaltet ein Fazit.<br>- Fazit enthält keine neuen Argumente. | Der Text beginnt mit einem klaren und einprägsamen Einleitungssatz, der das Thema der<br>Stellungnahme einführt und das Interesse der Leser:innen weckt.<br>2. Argumente:<br>- Mindestens ein Argument dafür:<br>Die Schüler:in nennt ein überzeugendes Argument, das die Vorteile von Schuluniformen darstellt, z.B. die<br>Förderung von Gleichheit unter den Schüler:innen und die Reduzierung von Mobbing aufgrund von<br>Kleidung.<br>- Mindestens ein Argument dagegen:<br>Die Schüler:in führt ein nachvollziehbares Argument an, das gegen die Einführung von Schuluniformen<br>spricht, z.B. die Einschränkung der individuellen Ausdrucksmöglichkeit der Schüler:innen.<br>3. Stellungnahme:<br>Die Stellungnahme ist klar und eindeutig. Die Schüler:in positioniert sich für eine der beiden Seiten, und<br>diese Position ist konsistent mit den zuvor genannten Argumenten.<br>4. Ableitbarkeit der Position:<br>Die vertretene Position ist logisch aus den Argumenten ableitbar, sodass die Leser:innen nachvollziehen<br>können, wie die Argumentation zur Schlussfolgerung führt.<br>5. Logische Schlüssigkeit der Argumentation:<br>Die Argumentation ist logisch strukturiert und schlüssig, sodass die Leser:innen den Gedankengängen der<br>Schüler:innen problemlos folgen können.<br>6. Fazit:<br>Das Fazit fasst die wichtigsten Punkte der Argumentation zusammen, ohne neue Argumente einzuführen.<br>Es bietet einen abschließenden Gedanken zur Thematik und verstärkt die gewählte Position. |

Tabelle: Vergleich unserer Aufgabenstellung mit Erwartungshorizont mit Beispiel 2 aus dem Prompt. Gelb markierte Textpassagen sind wörtlich identisch oder unmittelbar sinnäquivalent.

## Grobe Bewertung des seit 15.12.2024 "verbesserten" Tools

Wir haben das nach Auskunft des Unternehmens seit dem 15.12.2024 verbesserte Korrekturtool einer groben ersten Überprüfung unterzogen, um ein Bild davon zu gewinnen, welche unserer aufgelistet Mängelpunkte weiterhin bestehen. Da unser Prüfszenario für Testreihe A in den Prompt als ausführliches Beispiel übernommen wurden, ist es möglich, dass das Tool bei der Wiederholung dieser Testreihe besser als im Durchschnitt (für andere/beliebige Aufgaben) abschneidet, weshalb unsere Befunde eher zu optimistisch sein könnten. Die Tests wurden am 21.02.2025 durchgeführt.

| Mangel | Beschreibung | Gravität laut Abschnitt IV | Status quo im "verbesserten" Tool seit 15.12.2024<br><br>Auf der Grundlage einer kursorischen Sichtung |
|---|---|---|---|
| 1.a | Zufälligkeit der vorgeschlagenen Gesamtnote | **Fatales Gebrauchshindernis**<br><br>erschwerte Erkennbarkeit des Mangels durch Nutzende | **Mangel besteht weiterhin**<br><br>Die Ausgabe einer empfohlenen Gesamtnote erfolgt nicht immer; wenn sie erfolgt, ist sie soweit erkennbar unverändert variierend. |
| 1.b | Zufälligkeit der inhaltlichen Bewertung und qualitativen | **Fatales Gebrauchshindernis**<br><br>erschwerte | **Mangel besteht weiterhin**<br><br>Scheint schlimmer geworden. |
| | Rückmeldung | Erkennbarkeit des Mangels durch Nutzende | |
| 2.a | Unzuverlässige Erkennung von Falschbehauptungen | **Erhebliches Gebrauchshindernis**<br><br>Unterminierung des Wertversprechens, Delegitimierung von Prüfung und Benotung | **Mangel besteht weiterhin** |
| 2.b | Unzuverlässige Erkennung von Nonsense und Arbeitsverweigerung | **Erhebliches Gebrauchshindernis**<br><br>Unterminierung des Wertversprechens, Delegitimierung von Prüfung und Benotung | **Evtl. geringfügig abgemildert**<br><br>Verfehlung des Themas (Text 8) wird jetzt konsistent erkannt (unklar, ob auf andere Aufgabenstellung verallgemeinerbar!)<br>Arbeitsverweigerung (Texte 3, 7) werden nicht erkannt. |
| 3/a/b | Unzuverlässige Umsetzung einzelner Bewertungskriterien – z.B. a) Textlänge, b) Erkennung KI-generierter Abgaben | **Erschwernis des Gebrauchs**<br><br>Irreführung in Bezug auf die Fähigkeiten des Tools | **Nicht getestet / keine Verbesserung erwartet**<br><br>Da Fobizz in der Hilfestellung zum Tool erläutert, dass das Tool KI-generierte Texte nicht erkennen und Wörter nicht zuverlässig zählen kann, haben wir das nicht erneut getestet. |
| 4.a | a. Uneinheitliche und zufällige Bezeichnung von Fehlerkategorien | **Erhebliches Gebrauchshindernis**<br><br>mangelnde Transparenz und Einheitlichkeit der Bewertungsmaßstäbe; Delegitimierung von Prüfung und Benotung | **Abgemildert**<br><br>Die Fehlerkategorien wirken jetzt weniger willkürlich und sind eindeutig konstanter. Ab und zu gibt es aber noch einzelne Abweichler. |
| 4.b/c | Inkonsistentes Feedback durch a) Angabe von Fehlern, die keine sind; b) Widersprüchliche Angabe der Gesamtnote | **Erschwernis des Gebrauchs**<br><br>Irreführung in Bezug auf die Fähigkeiten des Tools, Motivationshemmnis und Irreführung der Schüler:innen | **Abgemildert**<br><br>a) Die Fehlertabelle bleibt jetzt häufiger einfach leer, wenn keine offensichtlichen Fehler da sind, anstatt Fehler zu enthalten, die keine sind. |
| 5.a/b | Die Umsetzung des Feedbacks führt nicht zur Verbesserung | **Fatales Gebrauchshindernis**<br><br>Mangel didaktischer Fundierung; Delegitimierung von Prüfung und Benotung | **Nicht getestet** |
| 6 | Erreichbarkeit einer Bestbewertung nur durch Einsatz von ChatGPT (Täuschung) | **Fatales Gebrauchshindernis**<br><br>Mangel didaktischer Fundierung; Delegitimierung von Prüfung und Benotung | **Mangel besteht weiterhin**<br><br>Die KI-generierte Abgabe (Text 9) erzielte weiterhin als einzige die Bestnote in allen Kategorien. |